\documentclass[11pt]{article}
\usepackage{epsfig,graphicx,cite}

\input amssym.def
\input amssym.tex

\makeatletter
\@addtoreset{equation}{section}
\makeatother

\def\ben{\begin{equation}}
\def\een{\end{equation}}
\def\half{{1 \over 2}}
\def\bea{\begin{eqnarray}}
\def\eea{\end{eqnarray}}

\def\p{\partial}

\def\sst#1{{\scriptscriptstyle #1}}
\def\del{\partial}
\def\opt{{{\sst{\rm opt}}}}

\def\mathbb{\Bbb}
\def\be{\begin{equation}}
\def\ee{\end{equation}}
\def\nn{\nonumber}
\newcount\hour \newcount\minute
\hour=\time  \divide \hour by 60
\minute=\time
\def\p{\partial}
\def\mathbb{\Bbb}

\def\ft#1#2{{\textstyle{\frac{\scriptstyle #1}{\scriptstyle #2} } }}
\def\fft#1#2{{\frac{#1}{#2}}}

\newcount\hour \newcount\minute
\hour=\time  \divide \hour by 60
\minute=\time
\loop \ifnum \minute > 59 \advance \minute by -60 \repeat
\def\nowtwelve{\ifnum \hour<13 \number\hour:
                      \ifnum \minute<10 0\fi
                      \number\minute
                      \ifnum \hour<12 \ A.M.\else \ P.M.\fi
         \else \advance \hour by -12 \number\hour:
                      \ifnum \minute<10 0\fi
                      \number\minute \ P.M.\fi}
\def\nowtwentyfour{\ifnum \hour<10 0\fi
                \number\hour:
                \ifnum \minute<10 0\fi
                \number\minute}
\def\now{\nowtwelve}
\newcommand{\hoch}[1]{$\, ^{#1}$}

\newcommand{\auth}{\Large\bf{M. Cveti\v c\hoch{1,2,6}, 
G.W. Gibbons\hoch{2,3,5} and C.N. Pope\hoch{1,3,4,5}
}}

\begin{document}
\begin{flushright}
\hfill { UPR-1280-T\ \ \ MI-TH-1623
}\\
\end{flushright}

\begin{center}

{\LARGE{\bf Photon Spheres and Sonic Horizons in   Black Holes
from  Supergravity and Other Theories}}

\vspace{9pt}
\auth

{\small

\vspace{8pt}{\hoch{1} \it Department of Physics, Beijing Normal University, Beijing,
100875, 
China}

\vspace{0pt}{\hoch{2}\it Department of Physics and Astronomy,\\
University of Pennsylvania, Philadelphia, PA 19104, USA}

\vspace{0pt}{\hoch{3}\it DAMTP, Centre for Mathematical Sciences,\\
 Cambridge University, Wilberforce Road, Cambridge CB3 OWA, UK}

\vspace{0pt}{\hoch{4}\it Mitchell 
Institute for Fundamental
Physics and Astronomy,\\
Texas A\&M University, College Station, TX 77843-4242, USA
}

\vspace{0pt}{\hoch{5}\it Laboratoire de Math\'ematiques et Physique Th\'eorique, CNRS-UMR 7350, Universit\'e de Tours, Parc de Grandmont, 37200 Tours, France}

\vspace{0pt}{\hoch{6}\it Center for Applied Mathematics and Theoretical Physics,\\
University of Maribor, SI2000 Maribor, Slovenia}

}

\newcount\hour \newcount\minute
\hour=\time  \divide \hour by 60
\minute=\time
\loop \ifnum \minute > 59 \advance \minute by -60 \repeat
\def\nowtwelve{\ifnum \hour<13 \number\hour:
                      \ifnum \minute<10 0\fi
                      \number\minute
                      \ifnum \hour<12 \ A.M.\else \ P.M.\fi
         \else \advance \hour by -12 \number\hour:
                      \ifnum \minute<10 0\fi
                      \number\minute \ P.M.\fi}
\def\nowtwentyfour{\ifnum \hour<10 0\fi
                \number\hour:
                \ifnum \minute<10 0\fi
                \number\minute}
\def\now{\nowtwelve}


%
\vspace{0pt}

\underline{ABSTRACT}
\end{center}

We study closed photon orbits in
spherically-symmetric static solutions of  supergravity theories,
a Horndeski theory, and a theory of quintessence.
These orbits lie in what we shall call  a {\it photon sphere} ({\it anti-photon sphere}) if
the orbit is unstable (stable). 
 We show that in all the asymptotically flat solutions we examine
that admit a 
regular event horizon, and whose energy-momentum tensor satisfies
the strong energy condition,
there is one and only one photon sphere outside
the event horizon.  
We  give an example of a Horndeski theory  black hole
(whose energy-momentum
tensor violates the strong energy condition)  
whose metric admits both  a photon sphere
and an anti-photon sphere. 
The uniqueness and 
non-existence also holds for asymptotically 
anti-de Sitter solutions in gauged supergravity. The latter  also exhibit
the projective symmetry that was first discovered  for 
the Schwarzschild-de Sitter metrics: the unparameterised null geodesics are
the same as when the cosmological or gauge coupling constant vanishes.
We also study the closely related problem of accretion flows by perfect 
fluids in these metrics.  
For a radiation  fluid, Bondi's  sonic horizon coincides with
the photon sphere. For a general polytropic 
equation of state this is not the case. Finally we exhibit
counterexamples to a conjecture of Hod's.

\pagebreak
\tableofcontents

\addtocontents{toc}{\protect\setcounter{tocdepth}{2}}

\pagebreak
\section{Introduction}

By Fermat's principle,
the study  of null geodesics in a static $(d+1)$-dimensional 
spacetime with
 metric
\ben
ds^2_{d+1} = -e^{2U(x)} dt ^2 + g_{ij} dx^i dx^j dx^i dx^j  
\een
may be reduced to the study of  the geodesics
of  the spatial manifold equipped with the  
conformally rescaled ``optical metric'' 
\ben
ds _\opt^2 = \gamma_{ij}dx^idx^j = e^{-2U} g_{ij}dx^idx^j \,.
\label{optical1}
\een  
The optical metric encodes more physical information than
just the optical properties of the spacetime. As we shall show
later, it is relevant to stability questions and
to the existence of York-Hawking-Page type phase transitions.   
Much more is known and is accessible in the spherically symmetric
case than for a general metric, and that is the situation
we shall consider in this paper. A great many 
spherically symmetric static solutions of Einstein's equations
are known, including those describing
black holes.  In particular, in recent years there has been a great deal
of activity constructing exact  solutions of the supergravity and related
equations of motion for spatial dimensions $d=3$ and higher. 
Since their stress-energy tensors, at least
without cosmological terms, satisfy the weak, dominant and   strong
energy conditions, one is assured that the properties
of such solutions are not artefacts of the matter content's 
being un-physical.

The motivations for our study include:
\begin{itemize}

\item
In the spherically symmetric case it is well known 
that unstable circular null geodesics are possible,
and that these circular null geodesics lie on a 
``photon sphere.''  In  principle
``anti-photon spheres,''
are also   possible\footnote{At an early stage of the  work 
reported here we 
were accustomed to refer to a sphere of stable  geodesics as a 
``whispering gallery.'' However the analogy with  more mundane
whispering galleries is not that close. As pointed out to us by Claude Warnick,
the term  ``whispering gallery'' is more appropriately  applied to 
the conformal boundary of anti-de Sitter spacetime.}. In such cases, the 
circular null geodesics
are stable.  These are much less familiar, and
to our knowledge there are no known asymptotically-flat examples 
that are regular outside a regular event horizon
and with matter content satisfying all of the three energy conditions
mentioned above.  Examples are known, however, in cases where naked
singularities are present \cite{Claudel:2000yi}.  
It has been suggested that the existence of
an anti-photon sphere is an indication that the solution may be unstable.
\cite{Keir:2014oka,Cardoso:2014sna}.

\item A less obvious aspect of photon spheres and anti-photon spheres
 is that they signal
the possibility of a York-Hawking-Page phase transition
\cite{York,Hawking:1982dh,Witten:1998qj}. This occurs
because the  Dirichlet boundary-value
problem in Euclidean  quantum gravity may have multiple
solutions that jump in number when the boundary passes through a 
photon sphere or an anti-photon sphere  \cite{Akbar}.

\item A number of conjectures have been made 
about photon spheres, and it is of interest to check them 
against our examples. In particular, we found that a conjecture of  
Hod \cite{Hod2} concerning a lower bound on the 
optical radius of a photon sphere, is violated  for 
dilatonic black holes with the dilaton coupling $a^2> 1$ and for
STU black holes with fewer than then two charges turned on.  On the other 
hand, a theorem of  Hod \cite{Hod}, concerning an upper bound on 
area-coordinate radius of a photon sphere,  is confirmed 
both for the STU black holes and dilatonic black holes.

\item It has been known for some time that
the existence of photon spheres 
affects the optical appearance of collapsing 
stars \cite{AT}, and  gives rise to shadows \cite{Synge}.
It is also known that the  optical radius governs the
high-frequency behaviour of the photon absorption cross
section,  and the high frequency
spectrum of quasi-normal modes \cite{Press,Goebel}.   

\item While the optical metric governs
the behaviour of null geodesics  parameterised 
by optical length, it may happen 
that two different metrics admit the same
unparameterised  null geodesics. This ``projective
equivalence'' actually  occurs for the Schwarzschild-de Sitter
or Kottler metric. The   unparameterised  null geodesics
are independent of the cosmological constant \cite{Islam,GWW,Casey:2011dw}.
Surprisingly, we find that this phenomenon is 
a rather general feature of the solutions we study.       

\item In the spherically symmetric case, 
each geodesic lies in a reflection-symmetric 
equatorial surface. 
The behaviour of the geodesics is heavily influenced
by the sign of the Gauss curvature of this surface
\cite{Gibbons:1993cy,GW,Gibbons:2008hb}, and in the 
asymptotically-flat case this allows a 
rapid evaluation of the light deflection at large
impact parameter \cite{Gibbons:1993cy,GW}. 
The  Gauss curvature also determines the shape, and indeed
the very possibility,
of isometrically embedding  the surface into Euclidean space
as a surface of revolution so as to provide an analogue model
of black holes  \cite{Carter,Cvetic:2012vg}. 
There is a close connection  between
the sign of the Gauss curvature and the existence  of
photon  and anti-photon spheres.

\item In the spherically-symmetric case
the steady  radial accretion or emission of  a test  perfect fluid 
must make a transition from sub-sonic to super-sonic
flow through a so-called  Bondi surface 
\cite{Carter}. For a 
radiation fluid for which the pressure is one third of the energy
density, the Bondi surface and  photon surface coincide.
As we shall show in an appendix, for an equation of state of the 
form $P= w \rho$ where $w$ is
a constant, the Bondi radius is located at a stationary point
of $ \frac{(-g_{tt}(R) )^{p-1}}{R^2}$, where $w= \frac{1}{2p-1}$.
If $p=2$ then $w=3$  and this gives the same condition for the existence   
of a photon sphere.

 \end{itemize} 

The plan of the paper is as follows.

In \S 2 we review in outline the general theory of
the optical metric of a static spherically symmetric spacetime
and its applications. Much, but not all,  of this
can be found scattered  in the existing literature but we thought it helpful
to assemble in one place and we have used this opportunity to establish
our notation. In particular appears to be  no consensus 
one what to call what we shall refer to as photon sphere and anti-photon sphere
and so we have spelled out in detail the usage adopted here. 

In \S 3 we discuss in detail the static spherically symmetric solutions 
of four-dimensional gauged and ungauged STU supergravity theory. 
After giving the metrics in 
a standard radial coordinate $r$  we introduce, in the
ungauged case,  an isotropic  coordinate $\rho$
which allows us to assign them an effective refractive index $n(\rho)$.
In the non-extremal case, when there is an event horizon, we are able to 
locate their unique photon sphere  and establish that
that  its location in the coordinate $r$ does not depend upon  the
gauge coupling constant.  We also verify that for non-extremal black holes 
the a theorem  of Hod's \cite{Hod} is satisfied, while Hod's 
conjecture \cite{Hod2} is violated if fewer than two charges are turned on.  
In the ultra-extremal case, 
which has naked singularities, we found that for a range of charges
there is both a photon sphere and an anti-photon sphere. 
We then investigate, by introducing an appropriate Binet type coordinate
$u$, analogous to that used in the central orbit problems 
of elementary non-relativistic dynamics, that  the projective properties
of the optical metric, i.e.\ its unparameterised
geodesics, do not depend on the gauge coupling constant $g$.
This result is confirmed at a more covariant  level by   
calculating the  Weyl projective tensor and finding it to be
independent of $g$. We conclude \S 3 by showing that similar results
hold for a class of dyonic solutions of 
gauged supergravity theories.           

In \S 4 we extend these results to static spherically  symmetric
solutions of  Einstein-Maxwell-Dilaton theory in four spacetime dimensions,
which depend upon an arbitrary Maxwell-dilaton coupling constant $a$.
These theories may be thought of as having a spacetime-dependent 
electric permittivity $\epsilon = \exp(-2a \phi)$, where
$\phi$ is the dilaton field, while preserving local Lorentz invariance.
These solutions permit a check that the conjecture of Hod in  \cite{Hod2} 
is violated for $a^2>1$, while Hod's theorem \cite{Hod} is obeyed.

In \S 5 we consider the static spherically symmetric solutions
of a particularly simple  Horndeski theory in which a metric $g_{\mu \nu}$
is coupled to a scalar $\chi$. For certain  values
of the constants entering the solution we find that the optical geometry
of the  metric $g_{\mu \nu}$ admits both a photon sphere and an anti-photon
sphere outside its Killing horizon. 
   
In \S 6 we treat a class of quintessence black holes due to Kiselev. 
Thy admit both a black hole horizon and  an analogue of the 
cosmological horizon that 
occurs in de Sitter spacetime. We find that, just as in the case
of de Sitter black holes, there  is just a  single photon 
sphere between the two horizons.

In \S 7 we provide a brief discussion of some static hyper-spherically 
symmetric solutions of gauged supergravity theories in five and seven 
spacetime dimensions. As in four spacetime dimensions 
we find at most a single photon hyper-sphere whose location is
independent of the gauged coupling constant $g$.

Finally in an appendix  we outline  a formalism for  irrotational
perfect fluids using a velocity potential   $\psi$, which 
may be regarded as  $k$-essence. Using this we are able to give
a novel treatment of accretion onto black holes, 
and to use it to locate the sonic or Bondi radius, which is the 
acoustic analogue of a photon surface. 

\section{General Theory}

\subsection{Notation and basic formulae}

In what follows we shall find it convenient to express the optical metric
in terms of various different radial
variables. We shall use $r$ for a generic radial variable, but 
reserve $r_\star$
for the radial optical distance or Regge-Wheeler tortoise coordinate,
and $R_\opt$, with $C_\opt = 2 \pi R_\opt $,   such 
that the optical metric (\ref{optical1}) becomes
\ben
ds^2_\opt = d r_\star^2 + R^2_\opt d \Omega ^2_{d-1} \label{ometric} 
\,,\een      
where $ d \Omega ^2_{d-1}$ is the unit metric on $\mathbb{S}^{d-1}$.   
Therefore  
\ben
R_\opt = e^{-U} R \,,
\een
where  $R$ is the ``area distance,'' such that 
the area of a 2-sphere measured in the physical
spacetime metric is  $4 \pi R^2$. Restricting (\ref{ometric})
to an equatorial 2-surface gives
\ben
ds^2_\opt|  = dr_\star^2 + R^2_\opt d \phi ^2 
\,,\qquad 0\le \phi < 2\pi  \,.
\een  
The Gauss   curvature is 
\ben
K _\opt = - \frac{1}{R_\opt } \frac{d^2 R_\opt}{d r_\star^2 }
\label{Gauss} \,.
\een
Any spherically symmetric metric is conformally flat, and 
so one may also  introduce an isotropic coordinate $\rho$  such that 
\ben
ds^2_\opt = n^2(\rho) \Bigl ( d \rho^2 + \rho^2 d \Omega^2_{d-1} \Bigr )\,.
\een   

The quantity $n$ may be interpreted in the language of 
elementary optics in Euclidean space  as  the refractive index
or slowness, so that the  ``speed  of light''   
$v= \ft{d \rho }{dt}$  in the coordinates $(t, \rho)$ is given by
$v= \frac{1}{n} \,.$   Thus we have 
\ben
R_\opt = n \rho = \frac{\rho}{v} \,. \label{oradius}
\een 
If $d w = - \frac{dr_\star} {R_\opt^2} = - \frac{d\rho}{n \rho ^2}$,
then, as we shall see in detail later, unparameterised geodesics of the optical 
metric satisfy an equation similar to Binet's equation for central orbits,
\ben
\frac{d^2 w}{d\phi^2} = - \frac{1}{2 }
 \frac{d }{d w} \frac{1}{R_\opt^2} 
\,.\label{Binet}
\een
Circular geodesics therefore correspond to extremals of the 
optical circumference at points
$r=\bar r$, i.e. for which   
\ben
R^\prime _\opt \big |_{r=\bar r} =0\,. \label{circular}
\een 
We have an unstable photon sphere if 
$R^{\prime \prime }_\opt\bigr |_{r=\bar r} >0$, and a stable photon
sphere, where light propagation is analogous to the acoustic
propagation in a SOFAR channel, 
if $R^{\prime \prime }_\opt\bigr |_{ r=\bar r} <0$.\footnote{A SOFAR 
(Sound Fixing and Ranging) channel arises in  a horizontal 
layer in the ocean where
the speed of sound attains a local maximum.  This acts like an
acoustic  waveguide, in
which low-frequency sound waves can travel large distances with little 
attenuation \cite{ewiwur,brek,Morawetz}.}

In the case of an asymptotically-flat black hole, 
$R_\opt$ goes to infinity both at infinity
and also at a regular horizon, and so there is always at least one 
photon sphere.
In general one might expect that there should be one more minimum  than
there are  maxima, that the 
outer and inner extrema should be minima, and that the inner extrema to have
$k$  maxima alternating with $k-1$  minima, there being $2k+1$  
extrema in all.   
From (\ref{oradius}) it follows that (\ref{circular}) is equivalent to
\ben
\frac{dv}{d\rho} = \frac{v}{\rho} \,. \label{graph} 
\een
Thus if we plot  the speed  $v=\frac{1}{n}$ against $\rho$
then photon and anti-photon  spheres  correspond to points on the graph
at which a straight line through the origin is tangent to it.
If the straight line touches the graph from above  we have
a photon sphere. If it touches it from below,  an anti-photon-sphere.
The slope at those  points $\rho=\bar\rho$ then equals the inverse optical
radius, $R_\opt(\bar\rho)^{-1}$.   

\subsection{Gauss curvature} 

In the usual case that there is just one photon sphere 
and the metric is asymptotically flat, we expect the graph
of $R_\opt(r_\star)$ to be convex, in which case
from (\ref{Gauss}) we see that the Gauss curvature
$K_\opt$ is negative. 
This allows a qualitative analysis
of the geodesics using the Gauss-Bonnet theorem 
\cite{Gibbons:1993cy,GW}. It also 
has implications for boundary rigidity
and the related inverse problem, which in turn
connects with holography and the AdS/CFT correspondence, as was 
observed in 
\cite{Porrati:2003na}.
Our situation relates to  what  geometers
call \emph{lens rigidity}, a  subject which also arises in connection with
invisibility cloaks, and related devices.   
The strongest general mathematical result
in this area  is directly applicable to the our present
work.  

\subsubsection{Lens Rigidity and Holography}

The basic idea is to idealise a static optical lensing device  
as a compact
connected $n$-dimensional  Riemannian manifold $\{M,g\}$ with a, 
not necessarily connected, boundary $\p M$, with     
light rays described as geodesics in the optical metric $g$.
If $\nu$ is the \emph{inward} pointing normal,
we define the $(2d-2)$-dimensional space $U^+\p M $ as 
the set of positions $x \in \p M$ and  inward pointing 
unit vectors $v$ 
such that $g(\nu,v) \ge 0$,  and   
$U^-\p M$ as the set of positions $x \in \p M$ 
and  outward  pointing unit vectors $v$ 
such that $g(\nu,v) \ne 0$. Then for all 
geodesics with initial tangent vector  
$v\in U^+\p M$ which after a finite  time $\tau>0$ first 
arrive at $\p M$, 
we get a map $S: U^+\p M \rightarrow U^-\p M$ called the
 \emph{scattering map}
or \emph{scattering data}.
Note that the scattering map is not defined if $\tau=\infty$,
in which case we say that the geodesic is \emph{trapped}  and 
may be defined as the identity if $\tau=0$. 
The scattering map  $S$ and the
time function $\tau: U^+ \p M \rightarrow {\Bbb R}_+ $  
are referred to as the \emph{lens data}. There is an obvious 
notion of equivalence, under  diffeomorphism of the boundary,
 of the notions of scattering map and lens data. The optical device
is said to be \emph{scattering rigid}  or \emph{lens rigid}  if 
the scattering data  or lens data determine the Riemannian  
manifold $\{M,g\}$ 
up  diffeomorphism. The freedom to make such diffeomorphisms
is the essential principle behind the construction of optical  
cloaking devices. Lens rigidity, if it holds, is 
the statement that that is 
the \emph{only}   freedom.  

Various theorems have been proved that demonstrate lens rigidity
under the restrictive assumption that the Riemannian manifold $\{M,g\}$ 
is \emph{simple}; that is, the boundary $\p M$ is strictly convex 
and for all $x \in M$ the exponential map $ {\rm exp}_x: {\exp}_x^{-1} (M)
\rightarrow M$ is a diffeomorphism. However if trapping takes place, then the
simplicity assumption does not hold.   There are comparatively few results
in that case. Since   trapping typically takes place
for light rays around black holes, this is an important gap
if one wishes to apply these results to the optical metrics
of static spacetimes. However, recently an important advance has been made by
Croke \cite{Croke1} (see also \cite{Croke2}), 
who shows that lens rigidity holds if 
\begin{itemize} 
\item   $d=2$, 
\item topologically 
$M \equiv S^1 \times I$, where $I$ is the unit interval,
\item the boundary $\p M$ is convex,
\item the Gauss curvature $K$ of $M$ is negative.
\end{itemize}

\subsubsection{Isometric Embedding}

Near a horizon one has \cite{Gibbons:2008hb}  
$K_\opt = \kappa$, where $\kappa$ is the surface gravity
of the horizon, which is of course a constant over the horizon.  
This has consequences for  the  popular way of visualizing  the geometry 
of a  two-dimensional Riemannian manifold.  
This is  to isometrically embed the
metric into Euclidean space. If the metric is invariant
under a circle action, one may attempt to embed it as a surface 
of revolution. If the embedding is  
\ben   
\bigl(r_\star, \phi \bigr ) \rightarrow 
\bigl ( x,y,z \bigr ) =\bigl(R_\opt (r_\star)\cos \phi ,
R_\opt (r_\star)\sin \phi, z(r_\star) \bigr )  \,,
\een
then $z(r_\star) $ satisfies the ordinary differential
equation:
\ben
\Bigl( \frac{dz}{d r_\star} \Bigr)^2 = 
1- \Bigl( \frac{R_\opt }{d r_\star} \Bigr)^2 \,.
\een
A solution will exist as long as 
\ben
\Bigl( \frac{R_\opt}{d r_\star} \Bigr)^2 \le 1\,. \label{obstruction} 
\een
For the Schwarzschild solution, this will be true as long as \cite{Carter}
\ben
R \ge \frac{9}{8} M \,.
\een
In \cite{Cvetic:2012vg},  the obstruction (\ref{obstruction}) was 
 shown   to apply to analogue models of black holes
constructed from graphene sheets.  
In terms of the isotropic coordinate $\rho$ and the ray velocity
$v$,   (\ref{obstruction}) becomes 
\ben
\Big( 1- \frac{\rho}{v} \frac{d v}{ d\rho} \Big)^2 \le 1 \,.  
\een

\subsubsection{Energy conditions and monotonicity of redshift}

The \emph{weak energy condition} implies
\ben
T_{\hat t \hat t} \ge 0 \,.
\een
If  the weak energy condition holds, then 
the Misner-Sharp mass $M(R)$ is non-decreasing 
and  bounded above by the ADM mass $M_{ADM}= M(\infty)$:
\ben
M(R) \le M_{ADM} \,. 
\een

The \emph{dominant energy condition} implies that 
\ben
T_{\hat t \hat t} -|T_{\hat R \hat R} | \ge 0\,,
\een
which implies the weak energy condition, as well as 

\ben
T_{\hat t \hat t} +T_{\hat R \hat R}  \ge 0\,.
\een

The \emph{Strong  energy condition} implies
\ben
T_{\hat t \hat t} +T_{\hat R \hat R} +  
T_{\hat \theta \hat \theta } + T_{\hat \phi  \hat \phi} \ge 0 \,.
\een

The \emph{Positive radial pressure condition } implies
\ben
T_{\hat R \hat R} \ge 0 \,.
\een

The \emph{Positive or Negative trace conditions} are
\ben
T \ge 0 \,, \qquad {\rm or} \qquad T \le 0 \,, \qquad{\rm respectively}\,. 
\een
Any  static solution of the Einstein equations
coupled to scalars and vectors, and with 
non-positive potentials for the scalars and a negative cosmological term,
satisfies the negative trace condition.       

The $R_{\hat t \hat t}$ orthonormal Ricci-tensor component of the
$d$-dimensional static metric
\be
ds ^2 =-\Phi^2 dt ^2 + g_{ij}dx^idx^j\,,
\ee
where $\Phi$ and $g_{ij}$ are independent of $t$,
is given by
\be
R_{\hat t\hat t} = \Phi^{-1}\, \nabla_g^2 \Phi\,,
\ee
where $\nabla ^2_g$ is the Laplace-Beltrami operator for the
spatial metric $g_{ij}$.  From this, it follows that the Einstein
equations $R_{\mu\nu}-\ft12 R g_{\mu\nu}=8\pi T_{\mu\nu}$ imply
(generalising the $d=4$ result of ref.~\cite{Whittaker})
\be
\nabla_g ^2  \Phi = 
 \fft{ 8 \pi\Phi}{d-2}\,  
\Big[ \fft{d-4}{d-2}\, T_{\hat t\hat t} +
 \bigl (T_{\hat t \hat t}+ \sum_{i}  T_{\hat i \hat i } \bigr ) \Big] 
\,.
\ee
(As a check on signs, note that in the Newtonian limit, where  we ignore 
$T_{\hat i \hat i}$,
then $\Phi = e^{U} \approx 1+U +\dots $ where $U$ is the Newtonian potential
and we 
recover the Poisson equation.)

In the case of a four-dimensional metric  with spherical symmetry this gives
\be
\frac{1}{\sqrt{g}}   \frac{d}{dr} 
\Bigl( \sqrt{g}  g^{rr}\frac{d\Phi}{dr} \Bigr) 
=  4 \pi \Phi \bigl( T_{\hat t \hat t} + T_{\hat r \hat r} + 
T_{\hat \theta  \hat \theta }+
T_{\hat \phi  \hat \phi } \bigr ) \,, 
\ee
where $g=\det{g_{ij}}$.
Thus 
\be
\sqrt{g}g^{rr}\frac{d\Phi}{dr} = \frac{ \kappa  A_H }{4 \pi} +  
\int^r_{r_H} 4  \pi \Phi 
\bigl(T_{\hat t \hat t} + T_{\hat r \hat r} + T_{\hat \theta  \hat \theta }+
T_{\hat \phi  \hat \phi }       \bigr )  \sqrt{g}dr \,, 
\label{Smarr} \ee
where $A_H$ is the area and $\kappa$ the surface
gravity of the horizon. 
By the strong energy condition, the integral on the right-hand side
 is non-negative,
and hence $|g_{tt}|$ is monotonically increasing.  Note that
if there is a negative cosmological constant, the same conclusion,
{\it a fortiori}, follows.
If we take the limit of (\ref{Smarr}) as $r\rightarrow \infty$ 
we obtain a form of the Smarr formula.

\subsection{Hod's theorem and a conjecture}

   In this subsection, we shall shall mainly use the area coordinate
$R$ as the radial variable.  As in the earlier discussion, we shall 
denote with a bar the value of the radial coordinate that corresponds to
a stationary point of the optical radius; i.e.\ a photon sphere or 
anti-photon sphere.

We consider the static metric
\bea
ds ^2 &=& -e^{2\gamma(R)}\, \Big(1- \frac{2M(R)}{R}\Big)\,  dt ^2 
+  \frac{dR^2}{ (1- \frac{2M(R)}{r}  ) } +
 R^2 (d \theta ^2 + \sin ^2 \theta d \phi ^2)  \,,\nn\\
&=& -e^{2U}\, dt^2 +\frac{dR^2}{ (1- \frac{2M(R)}{r}  ) } +
 R^2 (d \theta ^2 + \sin ^2 \theta d \phi ^2)  \,,
\label{MSmetric}
 \eea
where $M(R)$ is the Misner-Sharp mass. It satisfies
\bea
\frac{dM}{dR} &=& 4 \pi  R^2 T_{\hat t \hat t} \,,\label{dMdr}\\
\frac{d\gamma}{dR} &=& 
4 \pi  R  \frac{\Bigl ( T_{\hat t \hat t} +T_{\hat R \hat R} \Bigr )
}
{ (1- \frac{2M(R)}{R}  )   }   \,,\label{dgamdr}\\
\frac{d ( R^4 T_{\hat R \hat R}) }{dR} &=& 
-\frac{F}{(1-2 \frac{M(R)}{R})}\, \bigl(T_{\hat t \hat t} +
            T_{\hat R \hat R}\bigr)  + R T\,,\label{R2T}
\eea
where
\ben
T=T^\mu_\mu= -T_{\hat t \hat t} +
T_{\hat R \hat R} +T_{\hat \theta  \hat \theta }+   
T_{\hat \phi\hat \phi} 
\een
and 
\ben
F = 3 M(R)-R + 4 \pi R^2   T_{\hat R \hat R} \,.\label{Fdef}
\een
In the case of an isotropic fluid we have
\ben
T_{\hat t \hat t}= \rho \,, \qquad T_{\hat R \hat R} = 
T_{\hat \theta \theta}= T_{\hat \phi \hat \phi} = P \,,
\een
where $\rho$ is the energy density and $P$ is the pressure.
Our equations then reduce to the Tolman-Oppenheimer-Volkov equations
\bea
\frac{dP}{dR} &=& - 
(\rho +P) \frac{ M(R) + 4 \pi R^3 P }{R^2 (1-\frac{2M(R)}{R} )} \label{I}\\
\frac{dU}{dR} & =&     
\frac{ M(R) + 4 \pi R^3 P }{R^2 \bigl (1-\frac{2M(R) \bigr )}{R}}  \label{II}
\eea
whence 
\ben
dU = - \frac{d P}{\rho + P} \,. \label{III}
\een

\subsubsection{Hod's photon sphere theorems}

In the coordinates we are using, the optical radius for the
metric (\ref{MSmetric}) is given by
\ben
R_\opt(R)= R e^{-U}= 
R\, e^{-\gamma}\, \Big(1- \frac{2M(R)}{R}\Big)^{-\ft12}  \,.\label{hhh}
\een
It follows from (\ref{dMdr})  and (\ref{dgamdr}) that
\ben\fft{d (R_\opt^{-2})}{dR} = \fft{2}{R^4}\, e^{2\gamma}\, F\,,
\een
where $F$ is defined in eqn (\ref{Fdef}).
At either a  photon sphere or an  
anti-photon sphere,  
we have  $\frac{d R_\opt}{dR}=0$
and hence
\ben
F=0 \,,\qquad \Rightarrow \qquad \bar R = 3M(\bar R)+ 
4 \pi \bar R^3 \,T_{\hat R \hat R}(\bar R)\,.\label{optcon} 
\een

It is perhaps worth remarking that for an isotropic medium
for which the Tolman-Oppenheimer-Volkov equations hold, eqn  (\ref{optcon})
follows directly from (\ref{II}), by noting from $R_\opt= R e^{-U}$ 
that $dR_\opt/dR=0$ implies
\ben
\frac{1}{\bar R} = \frac{dU}{dR}\Big|_{R=\bar R}\,.
\een
  
Returning to  the general non-isotropic case, and  considering  a 
black hole, then
at the horizon $R=R_H$ the component  $ T_{\hat R \hat R}$ 
of the energy-momentum tensor vanishes, 
\ben
T_{\hat R \hat R}(R_H)  =0 \,,
\een 
and $R_\opt$ blows up:
\ben
\lim_{R \downarrow R_H} R_\opt(R) = \infty \,.
\een
Thus \emph{ $F$ is negative near the horizon} \cite{Hod}. 
On the other hand \emph{$F$ is positive near infinity}. 
Thus there must be at least one value of $R=\bar R$  for which  
$F(\bar R) =0$. Moreover the smallest  such value,
$\bar R_{\rm min}$, must be a local minimum, which corresponds to an
unstable photon sphere rather than a stable anti-photon sphere.
Thus $F$ is
negative for 
$R_H < R < \bar R_{\rm min}$.
Now if we assume the negative trace condition, 
it follows from  (\ref{R2T}) that
\ben
T_{\hat R \hat R}(\bar R_{\rm min}) <0 \,,
\een
and hence from (\ref{optcon}), we have
\ben
R_H < \bar R_{\rm min}  \le 3 M( \bar R_{\rm min})  
\le 3 M_{ADM}\,.\label{ineqs}
\een 
In particular, this implies Hod's theorem \cite{Hod}, namely, that
provided the trace of the energy-momentum tensor is negative, and that
the dominant energy condition holds, then
\be 
\bar R_{\rm min} \le 3 M_{ADM}\,. \label{hodtheorem}
\ee
A generalisation of (\ref{hodtheorem}) to higher dimensions has been 
given in \cite{gallo}.

   A further inequality proved by Hod in \cite{Hod} is as follows.  
Assuming the dominant energy condition, it follows from (\ref{dgamdr}) 
that $d\gamma/dR\ge0$, and hence, since $\gamma=0$ at infinity,
 $\gamma\le 0$. Thus, from (\ref{hhh}), we have
\be
R_\opt \ge R\, \Big(1 - \fft{2M(R)}{R}\Big)^{-1/2}\,.
\ee
From (\ref{ineqs}) we have $\bar R_{\rm min} \le 3M (\bar R_{\rm min})$, and
hence 
\be
R_\opt(\bar R) \ge \sqrt3 \bar R\,.\label{hodth2}
\ee

   The question of whether the closed photon orbit is stable or
unstable is governed by the sign of the second derivative of $R_\opt$ at
the radius of the orbit.  Using (\ref{dMdr}), (\ref{dgamdr}) and 
(\ref{R2T}), it follows, after imposing the condition (\ref{optcon}) 
that determines the orbital radius, that on the orbit we shall have
\be
\fft{d^2 R_\opt^{-2}}{dR^2}= \fft{2 e^{2\gamma}}{R^4}\, F'\,,
\ee
with
\be
F' \equiv \fft{dF}{dR}= 
-1 + 4\pi R^2\, (2 T_{\hat t\hat t} + T_{\hat\theta \hat\theta}
   +T_{\hat\phi\hat\phi})\,,\label{Fprime}
\ee
which is to be evaluated at the photon radius $R=\bar R$.  
The orbit is unstable
(a photon sphere) if $F'$ is negative, and stable (an anti-photon sphere)
if $F'$ is positive.

   As we show in later sections, in the case of theories such as 
supergravities, where the energy-momentum tensors satisfy all the relevant
energy conditions, we find that there is always exactly one closed
photon orbit outside the horizon of a regular black hole, 
and it is always unstable, corresponding
to a photon sphere.  However, it does not appear to be obvious
on general grounds from (\ref{Fprime}) that the energy conditions
are in themselves sufficient to guarantee the negativity of $F'$ 
at the photon orbit.  We show also that in the
case of ultra-extremal black holes (where there is a naked singularity),
there can be more than one photon orbit, with stable as well as unstable 
orbits.  We also study other examples with more 
exotic matter that does not obey all the usual energy conditions, and we 
show that in such cases there can exist multiple photon orbits outside
an horizon.

\subsubsection{Hod's conjecture }

Hod \cite{Hod2,Hod1} has made
some conjectures about photon surfaces in spherically-symmetric
geometries, and circular null geodesics in stationary
spacetimes. A special case of a conjecture in   \cite{Hod2} is  
that the optical radius $R_\opt$ of a 
photon surface in an asymptotically flat 
spacetime  with ADM mass $M_{ADM}$ satisfies
\ben
 R_\opt \ge 2M_{ADM}  \,. \label{Hod}
\een

   Both of Hod's theorems (\ref{hodtheorem}) and (\ref{hodth2}), and  the
conjecture (\ref{Hod}) may  be tested by the methods of this paper.
Unsurprisingly, the theorems hold in all the examples satisfying the 
assumptions under which they were derived.
We find that the conjecture (\ref{Hod}) 
is in fact violated in some circumstances. 
As we shall discuss later, we find that in the four-charge black holes 
of four-dimensional STU supergravity, the conjecture holds 
for the case where all the charges are equal 
(Reissner-Nordstr\"om), and for  pairwise equal charges
(string theory case, $a^2=1$).  However, the
conjectured inequality (\ref{Hod}) is not obeyed in the case where
only one charge is non-vanishing 
(Kaluza-Klein, $a=3$).  In section 4 we show that it is violated also in 
Einstein-Maxwell-Dilaton theories with $a^2> 1$.

\subsection{Geodesics and projective symmetry}

  The geodesics of the optical metric have two constants
of the motion:
\bea
{\rm Angular \, momentum}  \qquad R_\opt^2  \frac{d \phi}{ds_\opt } 
&=& h  \,.\\
{\rm Energy} \qquad \bigl (\frac{d r_\star}{ds_\opt}\bigr)^2 +
 R_\opt^2 \bigl ( \frac{d \phi} {d s_\opt } \bigr )^2 &=&1\,,
\eea
whence
\ben
\bigl(\frac{d r_\star  }{R_\opt^2  d\phi}\bigr)^2 +
 \frac{1} {R_\opt^2}=\frac{1}{h^2} = 
\bigl(\frac{dw}{d \phi} \bigr )^2  +   \frac{1} {R_\opt^2} 
\label{integral} \,. 
\een
If one differentiates (\ref{integral}) with respect to $w$ one obtains
the Binet type equation (\ref{Binet}).

An alternative procedure is to adopt isotropic coordinates, in which case
the geodesic equations may be cast into the standard form
for a central orbit problem.  Thus we make the standard redefinition 
$u=\frac{1}{\rho}$,  
and find that (\ref{integral}) becomes
\ben
\bigl( \frac{du}{d \phi} \bigr )^2 + u^2 = \frac{n^2}{h^2} \,,
\een
so that
\ben
\frac{d^2 u} {d \phi ^2} + u = \frac{P}{h^2u^2} \label{BINET} 
\een
with 
\ben
P = -\half \frac{\p n^2}{\p\rho}  \,,
\een
and where  $P$ is the acceleration of  the particle towards
the origin.  Equation (\ref{BINET}) is the standard form
of Binet's equation for central orbits.

\subsubsection{Projective symmetry} 

Differentiating (\ref{integral}) 
with respect to $u$  yields (\ref{Binet}), from which
it follows that  two metrics for 
which  $\frac{1}{R_\opt^2}$  differ
by a constant have the same unparameterised geodesics
and are thus  projectively equivalent, as explained in \cite{GWW}
where it was shown that the Weyl projective tensors
of  two such optical metrics are the same. 

A projective symmetry of this type
was first noticed for the Kottler metric, but not in this
language in \cite{Islam}. We shall see later that, rather remarkably,
all the gauged supergravity models that we study admit a projective
symmetry of this type.

\subsubsection{Shadows}
 
  For any curve, the angle $\delta$ made with the radial direction
satisfies 
\ben
\cot  \delta = \frac{1}{R_\opt }  \frac{d r_\star}{d \phi }   
\,.\een
For a geodesic it follows from (\ref{integral}) that
\ben
\sin \delta = \frac{h}{R_\opt } = \frac{h}{n\rho}\,,
\een 
which may be recognised as   Snell's law for a radially-stratified medium. 

For a geodesic that spirals around a photon sphere
or an anti-photon sphere  we have from (\ref{integral}) that
$h=R_\opt(\bar r)$, whence for such geodesics
\ben
\sin \delta(r) = \frac{R_\opt(\bar r)}{R_\opt(r)} \,. 
\label{shadow} 
\een
If $r > \bar r_{\sst{\rm max}}$, where $\bar r_{\sst{\rm max}}$ 
is the position of the
outermost photon sphere, then (\ref{shadow}) 
gives  the  angle  subtended 
by the shadow cast by this photon sphere \cite{Synge}.
For the Kottler metric
one has  
 \ben
\sin\delta =  \frac{3M}{R} 
\frac{ \sqrt{1 - \frac{2M}{R}   - \frac{1}{3} \Lambda R^2   }} 
{\sqrt{\frac{1}{3}  - 3 \Lambda M^2}}
\,, 
\een
and so $\delta= \frac{\pi}{2}$ at $R=3M$ (the photon sphere),
independently of $\Lambda$
as expected. However  the variation of $\delta$ with radius
definitely does depend upon $\Lambda$, since it is not a projectively 
invariant observable \cite{GWW,Gibbons:2010kv}. 

\subsubsection{Cross sections and quasi-normal modes}

If the metric is asymptotically flat
then $R_\opt(\bar r_{\sst{\rm max}})$ is the critical impact parameter  
such that light rays with smaller impact parameter cannot return to infinity.
Thus the high-energy limit of the absorption cross section is given by
\ben
\sigma = \pi  R^2_\opt(\bar r_{\sst{\rm max}}) \,. 
\een
For the Schwarzschild solution, the photon sphere is at $R=3M$ and thus 
\ben
 R_\opt(\bar r_{\sst{\rm max}}) = \sqrt{27} M \,,
\qquad \sigma =27 \pi M^2 \,. 
\een

Modes of oscillation of fields around black holes
can become trapped near photon spheres, and give rise to 
long-lived quasi-normal modes \cite{Press}.
Following \cite{Goebel}, one may estimate that in the large $l$ limit,
the real part
of the frequency behaves like     
\ben
\omega  \approx  \frac{l+\half} 
{R_\opt(\bar r_{\sst{\rm min}})} \,.
\een

\subsection{York-Hawking-Page phase transition}

    We conclude this brief review of the physics of photon spheres 
by noting its connection with the York-Hawking-Page
phase transition. The  York-Hawking-Page phase transition 
\cite{York,Hawking:1982dh,Witten:1998qj} 
plays a role when
we wish to count solutions of the Dirichlet problem
for the Riemannian Einstein equations \cite{York,Akbar}. The
geometries must be matched properly at the boundary.
Thus, in the spherically  symmetric case  we must match the circumference
 $C_\beta $ (or the local inverse temperature) of the $U(1)$ thermal  
circle, and
the circumference  $C_S$ of the boundary sphere
which we assume to be situated at $R=R_b$.
Now 
\ben
C_\beta= \frac{2 \pi}{\kappa} e^{U(R_b) }\,,
\een  
and 
\ben
C_S= 2 \pi R_b
\een
where $\kappa$, the surface gravity, 
is a function of the parameters defining the solution.
For example for the Kottler solution  
\ben
e^{2U} = (1-\frac{2M_{AD} }{R} + g^2 R^2 ) \,,
\een
where $M_{AD}$ is the Abbot-Deser mass, $g^2 = - \frac{\Lambda}{3}$,
and $\kappa = \kappa(M_{AD} ,g)$ is given by eliminating $r_H$ from the 
equations
\bea
\frac{M_{AD}}{R^2_H} + g^2 R_H  &=&  \kappa \\ 
1-\frac{2M_{AD}}{R_H} + g^2 R_H^2 &=& 0 \,.
\eea
Thus any saddle point of the path integral must satisfy 
\ben
\frac{\kappa C_S}{C_\beta} = R_\opt(R_b)   \label{intersection}\,,  
\een
where $R_\opt(R_b)$ is the optical radius of the boundary.
If we plot the graph of $R_\opt$ against $R_b$, 
the allowed values of $r_b$  correspond
to the intersection of the curve with the horizontal line
determined by the left-hand side of (\ref{intersection}).

As the left-hand side of  (\ref{intersection}) varies, solutions 
will appear or disappear in pairs, 
at values of $R_b$ for which 
\ben
\frac{dR_\opt}{d R_b} = 0\,.
\een        
\emph{That is, the number of solutions will jump
when the boundary is a  photon or an anti-photon sphere.}.  
   
Naively these  values correspond to phase transitions.
More accurately, they signal jumps in the minimum values 
of the Helmholtz free energy of the system.
\emph{It is a general feature that the location of the
boundary values $R_b$ for which the saddle points
jump in number is independent of the cosmological constant.}

\section{Static Spherically Symmetric STU Black Holes
 in Four Dimensions}

 In this section we shall explore in detail properties of 
photon spheres for static black holes in  four dimensional (gauged) 
supergravity theories. The  prototypes are black holes of 
maximally supersymmetric (gauged) supergravity theory, supported 
by four Abelian  gauge potentials and  three scalar axion-dilaton 
pairs. These fields in fact comprise a consistent truncation of the
maximal gauged supergravity to the $N=2$ supersymmetric gauged
STU supergravity theory.  Furthermore, since we are focusing solely 
on static solutions, only the three dilaton fields and the four 
electric gauge potentials are turned on. 

\subsection{Static four-charge STU black holes}{\label{fourcharge}}

For the static spherically-symmetric solutions  of the 
(maximally supersymmetric)
STU gauged supergravity  
the black-hole metrics are given by \cite{dufliu,CG}
 \ben
ds^2= -(H_1H_2H_3H_4) ^{-\half} f dt ^2 +
 (H_1H_2H_3H_4) ^{\half}  \Bigl [ {dr ^2 \over f} +
r^2  d \Omega ^2 _2   \Bigr ]\,, \label{STUmetric}
\een
with
\ben
f=1 -\frac{2m}{r} + g^2 r^2 H_1H_2H_3H_4\, , 
\een
and the harmonic functions $H_i$ are given by
\ben
H_i=1+\frac{q_i}{r}\, , \quad\quad i=1,2,3,4\,.
\een
The ADM mass and the  physical charges are  determined in terms of 
$m$ and $q_i$ as:
\ben
M_{ADM}=\sum_{i=1}^4 M_i\, , \quad \ M_i=\ft14(m+q_i)\, , \quad
Q_i^2=q_i(q_i+2m)\,, \quad i=1,2,3,4\,.\label{MQexp}
\een
For $m\ge 0$ and $q_i=2m\sinh^2\delta_i \ge 0$, the solutions have a 
regular horizon, and 
\ben
M_{ADM}= \ft14 m\,\sum_{i=1}^4(\sinh^2\delta_i+\cosh^2\delta_i)\ge 0\,, 
\quad Q_i=2m\sinh\delta_i\cosh\delta_i\ge 0\,.
\een
The solution can be uniquely parameterized in terms of physical 
charges $Q_i$, chosen, without loss of generality, to be positive, 
and  the positive ADM mass $M_{ADM}$, satisfying a BPS bound 
(of $N=8$ supergravity):
\ben
  M_{ADM}\ge \ft14 \sum_{i=1}^4 Q_i\, .
\een
We shall refer to these solutions as non-extremal ones.

If any of the $q_i\equiv -p_i$ parameters is chosen to be negative, 
the solution has a naked singularity at $r=p_{i\, max}$.  These solutions 
have mass below the BPS bound, and we shall refer to them as 
``ultra-extremal.''  Note from the expression for $Q_i^2$ in (\ref{MQexp})
that we must have $p_i\ge 2m $ in order that $Q_i$ be real.

\medskip
\noindent{\bf Isotropic coordinates and index of refraction}
\medskip
\\
In \cite{CGP}, the static non-extremal 
STU black holes  of the ungauged supergravity ($g^2=0$)  \cite{CYBPS,CYNON} were re-written
in terms of isotropic coordinates. 
Defining an isotropic radial  coordinate $\rho$ by
$r=\rho+m+ \ft{m^2}{4 \rho} $, it follows that 
\ben
\frac{dr^2}{1- \frac{2m}{r}}   +
r^2 \bigl( d \theta ^2 + \sin ^2 \theta d \phi ^2 \bigr )   = \Bigl(1 + \frac{m}{2\rho} \Bigr ) ^4
\Bigl \{  d \rho ^2 + \rho ^2 d\Omega^2   \Bigr \} \,.\label{isocoords}
\een
It now follows that 
\ben
(1+ \frac{m}{2\rho})^2   H_i=C_iD_i\,, 
\een
where $C_i$ and $D_i$ are spherically-symmetric harmonic functions:
\ben
C_i= 1 + \frac{me^{2 \delta _i} }{2\rho} \,,\qquad 
D_i= 1 + \frac{me^{-2 \delta _i}} {2\rho} \,.\label{CIDIsphsym}
\een
Note that $C_i$ and $D_i$, unlike the functions $H_i$ themselves, 
are harmonic in the
flat transverse 3-metric $d\rho^2+ \rho^2 d\Omega^2$.

   In terms of the isotropic radial coordinate, the metric (\ref{STUmetric})
becomes
\be
ds^2 = -\Pi^{-1/2}\, f_+^2\, f_-^2\, dt^2 + \Pi^{1/2}\, (d\rho^2 +
      \rho^2 d\Omega^2)\,, \label{ISOTROPICUNGAUGED}
\ee
where we have defined
\ben
\Pi=  \prod_{1\le I\le 4} C_i  D_i\,,\qquad f_\pm= 1 \pm \fft{m}{2\rho}\,.
\een
The scalar fields and gauge potentials can be written as
\be
X_i = \fft{\Pi^{1/4}}{C_i D_i}\,,\qquad A^i_\mu dx^\mu =
   \Big(-\fft{1}{C_i} + \fft{1}{D_i}\Big)\, dt\,.
\ee

Here we also provide the explicit parameterisation  for ultra-extremal 
solutions with one or more $q_i\equiv - p_i\le 0$. For $m>0$, the condition 
 $Q_i^2\ge 0$ on the charges implies that $p_i\ge 2m$.  The metric 
still takes the form (\ref{ISOTROPICUNGAUGED}), with the harmonic 
functions written as:
\ben
C_i=1+\frac{\alpha_i}{2\rho}\,  \quad D_i=1+\frac{\beta_i}{2\rho}\, , 
\een
where
\ben
\alpha_i=m+q_i +\sqrt{(m+q_i)^2 -m^2}\, , \quad \beta_i=m+q_i 
     -\sqrt{(m+q_i)^2 -m^2}\,.
\een
Note that these harmonic functions are well defined both for $q_i\ge 0$ (and 
reduce for $q_i=2m\sinh^2\delta_i$ to (\ref{CIDIsphsym})), as well as for 
$q_i\equiv -p_i$, as long as $p_i\ge 2m$. Again, the latter case 
corresponds to ultra-extremal solutions with a naked singularity 
at $\rho=-\frac{\beta_i}{2}$.  


     Note that the index of refraction is simply obtained from the 
form   (\ref{ISOTROPICUNGAUGED}) as:
\ben
n(\rho)=  \frac{\Pi^\frac{1}{2}}{f_+f_-}\,.  
\een 
While the index of refraction for non-extremal solutions blows-up at 
the outer horizon $\rho=\frac{m}{2}$, for  the ultra-extremal solutions 
it blows-up at the 
naked singularity $\rho= -\frac{\beta_i}{2}$.

\subsection{Photon spheres}
 
    In this subsection we analyse the properties of the 
photon spheres for these metrics. The radius of the photon sphere is 
simply determined  from (\ref{STUmetric})  as:
\ben
\frac{1}{R_\opt^{2}}  =  
\frac{1}{ r^2 H_1H_2H_3H_4}\,  
\Big(1 -\frac{2m}{r} \Big)+ g^2 \,. \label{Circum} 
\een
As argued in Section 2, one may note that the existence and location of any 
circular geodesic $r_{\star\, {\rm min}}$ or $r_{\star\, {\rm max}}$ 
is independent of $g^2$, but the optical radius
of any photon  $ R_\opt(r_{\star\, {\rm min}}) $  or anti-photon surface 
$R_\opt(r_{\star\, {\rm max}})$   will  depend upon  $g^2$, as do
the quasi-normal modes, and also the angle of any shadow. 
In the present case,  the 
optical circumference is an  extremum  when 
\ben
\frac{2(r -3 m)}{(r-2m) } = \sum _{i=1}^4 {q_i \over 
(r+q_i )}  \,.  \label{ext}
\een

\medskip
\noindent{\bf Non-extremal solutions}
\medskip
\\
The non-extremal solutions are parameterized by the positive quantity 
$m$ and the four positive quantities $q_i=2m\sinh^2\delta_i\ge 0$.  By 
analysing (\ref{ext}), it is 
straightforward to show that  outside the outer horizon at $r=r_+$,
there is only one extremum, which is located
at $r=\bar r>3m$.\footnote{For $g^2=0$, 
$r_+=2m$,  and  for $g^2>0$, $r_+<2m$.  and thus the result of the analysis above applies to both cases.
}
Namely,  the left-hand side of (\ref{ext}) is a monotonically-increasing 
function of $r$, with a negative pole at $r\to 2m^+$, zero at $r=3m$ and 
approaching $2$ as $r\to \infty$. On the other hand, the right-hand side 
is a monotonically-decreasing function of $r$, with a positive finite value 
at $r=2m$ and approaching $0$ as $r\to \infty$. Thus there is only {\it one} 
common solution in this domain,  at $r=\bar r > 3m$.
It is straightforward to show  that  the extremum is a {\sl minimum}, and
so it gives a single  unstable circular null geodesic.

In the following we shall also address the theorem (\ref{hodtheorem}) and  
the conjecture (\ref{Hod}) of Hod \cite{Hod,Hod2}.

We can also show that for the case of fewer than two charges turned on, the
 conjecture (\ref{Hod}) of Hod \cite{Hod2} is violated.
For concreteness we take only $q_1=q_2\ne 0$.  In this case we have the 
ratio
\ben 
\frac{R_\opt(\bar r)^2}{ 4M_{ADM}^2}= 
\frac{1}{16}\frac{\left(3+\sqrt{8{\tilde q}+9}+4\tilde{q}\right)^2
\left(3+\sqrt{8{\tilde q}+9}\right)}{\left(-1+\sqrt{8{\tilde q}+9}
\right)({\tilde q}+1)^2)}\ge 1\,,
\een
where ${\tilde q}\equiv \frac{q}{2m}$. The equality is attained  in the 
limit $\delta\to \infty$. 
 The analysis of the single charge case (e.g.\ only $q_1\ne 0$) 
reveals that the conjecture is violated when 
${\tilde q}_1\equiv \frac{q_1}{2m}
\ge 13.94$.

It is straightforward to show that Hod's theorem \cite{Hod}, given in 
(\ref{hodtheorem}),   
is satisfied. Namely, one can write
\ben
\bar R =\prod_{i=1}^4 (\bar r+q_i)^{\frac{1}{4}}\le 
\frac{1}{4}\sum_{i=1}^4(\bar r +q_i)= 3M_{ADM} + {\bar r}-3m-\frac{1}{2}\sum_{i=1}^4 q_i
 \le 3M_{ADM}\, .
\een
The first inequality above is due to the inequality of geometric and 
arithmetic means. The second inequality is due to the fact that:
\ben
{\bar r}-3m-\frac{1}{2}\sum_{i=1}^4 q_i=-\frac{1}{2}\sum_{i=1}^4\frac{q_i(q_i+2m)}{{\bar r}+q_i}\le 0\, .
\een
where where the first equality is due to (\ref{ext}).

  One can also show that
the inequality in Hod's theorem (\ref{hodth2}) is also satisfied.

\medskip
\medskip
\noindent{\bf Ultra-extremal solutions}
\medskip

 The occurrence of photon spheres in  extremal black holes
has been extensively studied, for example in \cite{prad,khoong}, and
we shall not consider this case further here.  Instead, we move on
to a study of the ultra-extremal case, where one or more of 
the $q_i$ parameters is 
negative. For $q_i\equiv -p_i$, with $p_i\ge 2m$ and $i=1,\cdots k$, the 
extremum equation for the photon radius takes the form:
\ben
\frac{2(r -3 m)}{(r-2m) } = \sum _{i=1}^k {-p_i \over 
(r-p_i )} +\sum _{j=k+1}^4 {q_j \over 
(r+q_j )}  \, .  \label{extultra}
\een
A straightforward analysis shows that a necessary condition for the 
above equation to have a solution is that $k=1$, i.e.\ only one 
of the $q_i$ is negative.  To see this, we take $q_1=-p_{max}$ and $k\ge 2$,   
so Eq. (\ref{extultra})
can be written as:
\ben
2+\frac{r\, (p_{max}-2m)}{(r-2m)(r-p_{max})} + \sum _{i=2}^k {p_i \over 
(r-p_i )} =  \sum _{j=k+1}^4 {q_j \over 
(r+q_j )}  \,,  \label{extultrap}
\een
where the $p_i$ for $i\ge2$ satisfy $p_i\le p_{max}$. The naked singularity
is located at $r=p_{max}$.  
The left-hand side of (\ref{extultrap})  
is manifestly larger than 2 for $r\ge p_{max}$.  
The necessary condition for the solution to exist is that  the 
right-hand side of (\ref{extultrap}) be $\ge 2$ for $r=p_{max}$. 
This condition  cannot be satisfied for $k\ge 2$, thus demonstrating that
photon spheres can arise for ultra-extremal black holes only if
just a single $q_i$ is negative. 

    For  $k=1$, 
the left-hand side of (\ref{extultrap}) lacks the final term, and 
it remains $\ge 2$ for $r\ge p_{max}$. In this case the necessary 
condition that the right-hand side be $\ge 2$ for $r=p_{max}$ reduces to 
the condition
\ben
\prod_{i=2}^4\frac{q_i}{p_{max}}\ge  \sum_{i=2}^4\frac{q_i}{p_{max}} +2\,,
\een
which can be satisfied for a range of parameters $q_i$. For  
the case $q_2=q_3=q_4\equiv q$, the above inequality  is satisfied for  
$q\ge 2p_{max}$.

Further focusing on the latter case, namely,   
$q_1\equiv -p$ and  $q_2=q_3=q_4\equiv q$, eqn (\ref{extultrap}) becomes
\be
2 + \fft{\tilde r(\tilde p-1)}{(\tilde r-1)(\tilde r-\tilde p)} =
   \fft{3\tilde q}{(\tilde r + \tilde q)}\,,
\ee
where we have defined 
\be
{\tilde r}\equiv \frac{r}{2m}\,, \quad \quad {\tilde q}=\frac{q}{2m}\,, 
\quad \quad  {\tilde p}\equiv \frac{p}{2m} \,.
\ee
Plotting the left and right hand sides one can see that there
will be either two intersections or none, in the region $\tilde r>\tilde p$
outside the naked singularity, depending on the choice of the parameters.
The critical intermediate case occurs if the parameters are such that
the left and right hand sides, and also their first derivatives, are equal
for some $\tilde r_{crit}$.  These two conditions allow one to derive
the corresponding values of $\tilde p_{crit}$ and $\tilde r_{crit}$ in terms
of $\tilde q$.  The result is
\bea
{\tilde r}_{crit}&= &\textstyle{\frac{1}{2}}\left(4{\tilde q}+3-\sqrt{12{\tilde q}^2+12{\tilde q}+9}\right)\, , \\  \nn
{\tilde p}_{crit}&=& \frac{(26{\tilde q}^2+27{\tilde q}+9)\sqrt{12{\tilde q}^2
+12{\tilde q}+9}-90{\tilde q}^3-138{\tilde q}^2-99{\tilde q}-27}{(2{\tilde q}
+1)\sqrt{12{\tilde q}^2+12{\tilde q}+9}-6{\tilde q}^2-6{\tilde q}-3}\,. 
\label{crit}
\eea
It is straightforward to show that  $2m\le p_{crit} \le \ft12 q$, and  
$r_{crit}\ge  p_{crit}$, i.e., the extremum is located outside 
the naked singularity.

   In summary, we have shown that for  $p\ge p_{crit}$, 
Eq. (\ref{extultra}) has no solution, whilst for $p\le p_{crit}$, 
Eq. (\ref{extultra}) has two solutions.  In the latter case, the outer 
solution corresponds to a minimum, which is
stable (an anti-photon sphere) and the  
inner solution to a maximum, which is therefore unstable (a photon sphere).


\subsection{Projective symmetry for the general STU black holes}\label{projSTU}

 The optical metric of a static black hole can always be cast in the form
\be
\frac{du^2} {k^2(u)} + \frac{1}{k(u)} d \Omega_2 ^2 
\label{one}  \,.
\ee
It was shown in \cite{GWW} 
that the Weyl projective tensor
depends 
only on $k^\prime$ and $ k^{\prime \prime}$.  
For metrics  of the form (\ref{one}),
one can  assume that coordinates may be chosen so that any geodesic
lies  in the equatorial plane   $\theta=\frac{\pi}{2}$.
The geodesics then satisfy
\ben
\bigl (\frac{du}{d \phi} \bigr ) ^2 + k = \frac{1}{h^2}  
\label{Clairaut}\een
where $h$ is   Clairaut's constant, which may be thought of
as the angular momentum or impact parameter.
Differentiating (\ref{Clairaut}) we obtain the second-order
equation
\ben
\frac{d^2u}{d \phi^2} +\half k^\prime =0\,.
\een

The optical metric of the static STU black hole (\ref{STUmetric})  
can be cast in the form
  (\ref{one}), by introducing a coordinate $u=u(r)$ such that
\be
k(u)= \fft{f}{r^2\, H}\,,\qquad \fft{{u'}^2\, f^2}{H} =k^2(u)\,,
\label{kdef}
\ee
where $H\equiv \prod_{i=1}^4 H_i(r)$ and $H_i(r)$ and $f(r)$ are defined in Eq. (\ref{STUmetric}). This implies that $u$ is given by
\be
u =\int^r\fft{dr'}{\prod_i (r'+q_i)^{\ft12}}\,.
\ee
This integral can be evaluated, to give
\be
u = \fft{2}{(q_2-q_3)^{\ft12}\, (q_1-q_4)^{\ft12}}\,
  F\Big(\fft{(q_1-q_4)^{\ft12}\, (r+ q_2)^{\ft12}}{
             (q_2-q_4)^{\ft12}\, (r+q_1)^{\ft12}} ;
\fft{ (q_1-q_3)^{\ft12}\, (q_2-q_4)^{\ft12}}{
     (q_2-q_3)^{\ft12}\, (q_1-q_4)^{\ft12}}\Big)\,,\label{usol}
\ee
where the incomplete elliptic function of the first kind is defined by
\be
F(\sin\varphi; \kappa) =\int_0^\varphi 
\fft{d\theta}{\sqrt{1-\kappa^2 \,\sin^2\theta}}
\,.
\ee
Note that the function $k(u)$, defined by the first equation in
(\ref{kdef}),  is given by
\be
k(u) = \frac{1}{R^2_{\opt}}=\fft{1}{r^2\, H}\, 
\Big(1-\fft{2m}{r}\Big)  + g^2\,,
\ee
where $u$ is defined in terms of $r$ by (\ref{usol}), and thus
the projective symmetry condition is satisfied (since $k'(u)$ is
independent of $g^2$).  
The expression for $r$ in terms of $u$ can be made explicit in terms of the
Jacobi elliptic function sn$(v;{\tilde k})$, which is related to the
incomplete elliptic integral by $F(x;{\tilde k})=v$, 
where $x=$sn$(v;{\tilde k})$.  Thus we find 
\be
r= \fft{q_1(q_2-q_4)\, \hbox{sn}^2(v;{\tilde k}) -q_2(q_1-q_4)}{
    (q_1-q_4) - (q_2-q_4)\, \hbox{sn}^2(v;{\tilde k})}\,,
\ee
where
\be
v= \ft12 (q_2-q_3)^{\ft12}\, (q_1-q_4)^{\ft12}\, u\,,\qquad
{\tilde k} = \fft{ (q_1-q_3)^{\ft12}\, (q_2-q_4)^{\ft12}}{
     (q_2-q_3)^{\ft12}\, (q_1-q_4)^{\ft12}}\,.
\ee

  For the special case of pair-wise equal charges 
$q_1=q_3$ and $q_2=q_4$,  the transformation is invertible in terms 
of elementary functions:
\be 
u=\frac{1}{q_2-q_1}\log(\frac{r+q_2}{r+q_1})\,,
\ee
and
\be 
r=\frac{q_1x-q_2}{1-x}\, ,\quad \quad x=\exp((q_2-q_1)u)\,.
\ee
For the Reissner-Nordstr\"om case $q_1=q_2=q_3=q_4\equiv q$,  
the relation between $u$ and $r$ s very simple, namely
\be
u=\frac{1}{r+q}\,.
\ee
In this case $u= \frac{1}{R}$, where $R$ is the area distance.
It is easy to check that the geodesics of the optical metric
are given by Weierstrass functions of the azimuthal coordinate 
$\phi$  in this case (c.f. 
\cite{Villanueva:2013zta}). Setting $q=0$ and we recover the
Schwarzschild case \cite{Gibbons:2011rh}.

In general case one may define
$\tilde u = \frac{1}{r}$ and obtain the equation
\be
\bigl( \frac{d \tilde u}{d \phi} \bigr ) ^2 + \tilde u^2 -2m^3 +
(g^2-\frac{1}{h^2} ) H(\tilde u) =0\,. 
\ee
It follows that the geodesics of the optical metric
are  given in general  by Weierstrass functions of the azimuthal coordinate 
$\phi$.  

One may also evaluate the Weyl projective tensor 
directly in the $r$  coordinates  and verify that it does not depend on $g^2$.

\subsection{Dyonic  solutions of the gauged STU model}{\label{dyonicsol}

Here we show that analogous properties of the STU black holes also hold for 
the case of the dyonic black hole solutions found in \cite{lupapo}.  These
black holes are solutions of the theory described by the Lagrangian
\be
{\cal L}= \sqrt{-g}\Big[ R -\ft12 (\del\phi)^2 -\ft12 e^{-\sqrt3 \phi}\,
  F^2 + 6 g^2\, \cosh\Big(\ft1{\sqrt3}\phi\Big)\Big]\,.
\ee
This theory is a the bosonic sector of a consistent truncation of 
${\cal N}=8$ gauged supergravity in which just a single $U(1)$ gauge 
field is retained.  It is also a consistent truncation of gauged STU
supergravity.  The dyonic black hole solution is given by \cite{lupapo}
\be
ds^2 = -(H_1 H_2)^{-\ft12}\, f\, dt^2 + (H_1 H_2)^{\ft12} \Big(
  \fft{dr^2}{f} + r^2\,d\theta^2 + r^2\sin^2\theta\, d\varphi^2)\Big)\,,
\ee
where 
\bea
\phi&=&\fft{\sqrt3}{2}\log\fft{H_2}{H_1}\,,\qquad
f= f_0 + g^2 r^2\, H_1 H_2\,,\qquad f_0=1-\fft{2m}{r}\,,\nn\\
A&=& \sqrt2\Big( \fft{1-\beta_1 f_0)}{\sqrt{\beta_1\gamma_2}\,H_1}\, dt
+ 2m\gamma_2^{-1}\, \sqrt{\beta_2\gamma_1}\, \cos\theta\, d\varphi\Big)\,,\nn\\
H_1 &=& \gamma_1^{-1}\, (1-2\beta_1\,f_0+\beta_1\beta_2\,f_0^2)\, , \qquad 
H_2 = \gamma_2^{-1}\, (1-2\beta_2\,f_0+\beta_1\beta_2\,f_0^2)\,.
\eea
The constants $m$, $\beta_1$ and $\beta_2$ characterise the mass, electric and
magnetic charges \cite{lupapo}, and the constants 
$\gamma_1$ and $\gamma_2$ are given in
terms of $\beta_1$ and $\beta_2$ by
\be
\gamma_1= 1-2\beta_1 + \beta_1\beta_2\,,\qquad \gamma_2=1-2\beta_2 +
  \beta_1\beta_2\,.
\ee
The constants $\beta_1$ and $\beta_2$, which must each lie in the range
$0\le\beta_i\le 1$, are further constrained by the requirement, 
for positivity of the functions $H_i$, that $\gamma_i\ge0$. 

  The radius of the extremal photon sphere satisfies
\ben
\frac{2(r -3 m)}{(r-2m) } = 
-r\, \left(\frac{H^\prime_1}{H_1}+\frac{H^\prime_2}{H_2}\right) \,, 
\label{dyonic}
\een
where $H^\prime_i\equiv \frac{dH_i}{dr}$.
It is straightforward to show that
\ben 
-r\, H_1^\prime =  \frac{2\beta_1\, [1+x(1-\beta_{2})]}{(1+x)^2}
\ge 0\,,
\een
where $r=2m(1+x)$, with an analogous result for $H_2'$ in which the labels
1 and 2 are interchanged.  Since $r\ge 2m$ corresponds to $x\ge0$,
it is manifest that the right-hand side  of (\ref{dyonic}) 
is always non-negative for $x\ge0$ (i.e. $r\ge 2m$).  It approaches the value
$2(\beta_1+\beta_2)$ as $x$ goes to zero, and it goes to zero as $x$ goes
to infinity.

 Furthermore, one can see that the right-hand side of (\ref{dyonic}) 
is a monotonically decreasing function of x. Namely, one can show that
 \ben  \left(-r\, \frac{H^\prime_1}{H_1}\right)^\prime=
-\frac{\beta_1}{m\, H_1^2}
 \left[\gamma_1+\beta_{2}(1-\beta_{1})+2x\gamma_{1}+
   x^2\gamma_{1}(1-\beta_{2})\right]\le 0\,, 
 \een 
with an analogous result where the labels 1 and 2 are interchanged.
Thus there is only one solution of (\ref{dyonic}), at $r=\bar r \ge 3m$, just 
as in the 4-charge solution of section \ref{fourcharge}.

\section{Einstein-Maxwell-Dilaton Black Holes}

   In this section, we study the properties of photon spheres for static
black holes in the family of Einstein-Maxwell-Dilaton (EMD) theories.

\subsection{Static black holes in EMD theories}

   Einstein-Maxwell-Dilaton gravity is described by the Lagrangian 
\be
{\cal L}= \sqrt{-g}\, (R - 2(\partial\phi)^2 - e^{-2a\phi}\, F^2)\,.
\label{emdlag}
\ee
 The static black hole
solution is given by \cite{gibbmaed}
\bea
ds^2 &=& -\Delta \, dt^2 + \Delta^{-1}\, dr^2 + R^2\, d\Omega_2^2\,,\cr
e^{-2a\phi} &=& F_-^{\ft{2a^2}{(1+a^2)}}\,,\qquad A=Q\cos\theta\, d\varphi\,,\cr
\Delta &=& F_+\, F_-^{\ft{(1-a^2)}{(1+a^2)}}\,,\qquad
  R^2 = r^2 \, F_-^{\ft{2 a^2}{(1+a^2)}}\,,\cr
F_\pm &=& 1  - \fft{r_\pm}{r}\,, 
\label{gibmaesol}
\eea 
and
\ben
M_{ADM}= \half \Big( r_+ + \frac{1-a^2}{1+a^2} r_-\Big)\,,\qquad Q^2 = 
\frac{r_+r_-}{1+a^2}\,.
\een

If a potential of the type considered in  \cite{Gao:2004tu} is added,
namely
\ben
  V(\phi) =-\frac{2\lambda}{3(1+a^2)^2}\left[a^2(3a^2-1)e^{-\frac{2\phi}{a}}+(3-a^2)e^{2a\phi}+8a^2e^{\left(a\phi-\frac{\phi}{a}\right)}\right]
  \, , 
 \een
 the  only change to the solution
is in the function 
$\Delta$, which is then given by \cite{Gao:2004tu}
\ben
\Delta = F_+ F_-^{\frac{1-a^2}{1+a^2}} -\frac{\lambda}{3} R^2 \,.
\een

\subsubsection{Isotropic coordinates and refractive index}

If $\lambda =0$, we can introduce an isotropic radial coordinate $\rho$, 
defined by
\be
\log\rho = \int\fft{1}{r\, \sqrt{F_-\, F_+}}\, dr\,,
\ee
which implies that, with a convenient choice for the constant of integration,
\be
r= \rho\, \Big(1+\fft{u^2}{\rho}\Big)\Big(1+\fft{v^2}{\rho}\Big)\,,
\ee
where we have re-parameterised the constants $r_\pm$ in terms of 
constants $u$ and $v$ as
\be
r_+= (u+v)^2\,,\qquad r_-= (u-v)^2\,.
\ee
In terms of the new quantities, we have
\be
F_- = \fft{\Big(1+\fft{uv}{\rho}\Big)^2}{\Big(1+\fft{u^2}{\rho}\Big)
   \Big(1+\fft{v^2}{\rho}\Big)}\,,\qquad
F_+ = \fft{\Big(1-\fft{uv}{\rho}\Big)^2}{\Big(1+\fft{u^2}{\rho}\Big)
   \Big(1+\fft{v^2}{\rho}\Big)}\,.
\ee
The metric now takes the form
\be
ds^2=-\Delta\, dt^2 + \Phi^4\, (d\rho^2+\rho^2\, d\Omega_2^2)\,,
\ee
where
\be
\Phi^2 = \fft{R}{\rho}= \Big[ \Big(1+\fft{u^2}{\rho}\Big) 
   \Big(1+\fft{v^2}{\rho}\Big)\Big]^{\ft{1}{1+a^2}}\, 
  \Big(1+\fft{uv}{\rho}\Big)^{\ft{2a^2}{1+a^2}}\,,\label{Phiiso}
\ee
and with the dilaton given by
\be
e^{2a\phi}= \Big[ \Big(1+\fft{u^2}{\rho}\Big) 
   \Big(1+\fft{v^2}{\rho}\Big)\Big]^{-\ft{2a^2}{1+a^2}}\, 
  \Big(1+\fft{uv}{\rho}\Big)^{\ft{4a^2}{1+a^2}}\,.\label{diliso}
\ee

The effective refractive index $n(\rho)$ in this representation
is given by
\ben
n(\rho) = \frac{\Phi ^2(\rho) } {\sqrt \Delta(\rho)} = \frac{\left[(1+\frac{u^2}{r})(1+\frac{v^2}{r})\right]^{\frac{3}{1+a^2}}}{(1+\frac{uv}{r})^\frac{2}{1+a^2}(1-\frac{uv}{r})^{\frac{2(1-a^2)}{1+a^2}}}\, .
\een

\subsection{Photon spheres and Hod's conjecture}

  For the static dilatonic black holes solutions \cite{Gao:2004tu},  
discussed above,   the photon radius is of the form:
\ben
\frac{1}{R_\opt^{2}}  =\frac{1}{r^2} F_+ F_-^{\frac{1-3a^2}{1+a^2}}
 - \frac{1}{3} \lambda\,. \label{phr}
\een 
Thus the independence  of the location of the photon spheres
on the cosmological constant continues to hold in this case as well. 
%
%
The extremal values of the photon spheres are at values of $r=\bar r$
satisfying  the equation
\ben
\frac{3}{r} - \frac{1}{r-r_+} + \frac{3 a^2 - 1}{ 1+a^2}
\frac{r_-} {r } \frac{1}{r-r_-}   =0\,.
\een
This  quadratic equation determines two stationary points, $r=b_\pm$,
with
\ben
b_\pm = \fft14 \Big[3 r_+ + (2-x) r_- \pm\sqrt{[(2-x) r_- -r_+]^2+
       8r_+ (r_+ -r_-)}\Big]\,, \label{extremad}
\een
where $x\equiv (3a^2-1)/(a^2+1)$.  Noting that 
$1+x=4 a^2/(a^2+1)\ge 0$ and $3-x=4/(a^2+1)\ge0$, it follows
that $x$ lies in the range $-1\le x\le 3$.  Assuming $0\le r_-\le r_+$ we 
have
$\sqrt{ [(2-x) r_- - r_+]^2 +8 r_+ (r_+ - r_-)} \ge |Z|$, where we define
\ben
Z= (2-x) r_- - r_+
\een
(which may have either sign).
It then 
follows that 
\ben
b_+ - r_+  \ge \ft14(Z + |Z|) \ge0\,,\qquad 
b_- - r_+ \le \ft14(Z-|Z|) \le 0\,,
\een
and so the larger stationary point always lies outside the outer horizon,
while the smaller stationary point lies inside.

\subsubsection{Photon spheres: Non-extremal dilatonic solutions}

  In \cite{Hod}, Hod  conjectured the bound (\ref{Hod}) 
 for static black holes, or, in other words, 
\ben
{\cal N}\equiv  \frac{R_\opt(\bar r)^2}{4M_{ADM}^2}\ge 1\, ,
\een

\begin{figure}[ht]
\begin{center}
\includegraphics[height=8cm]{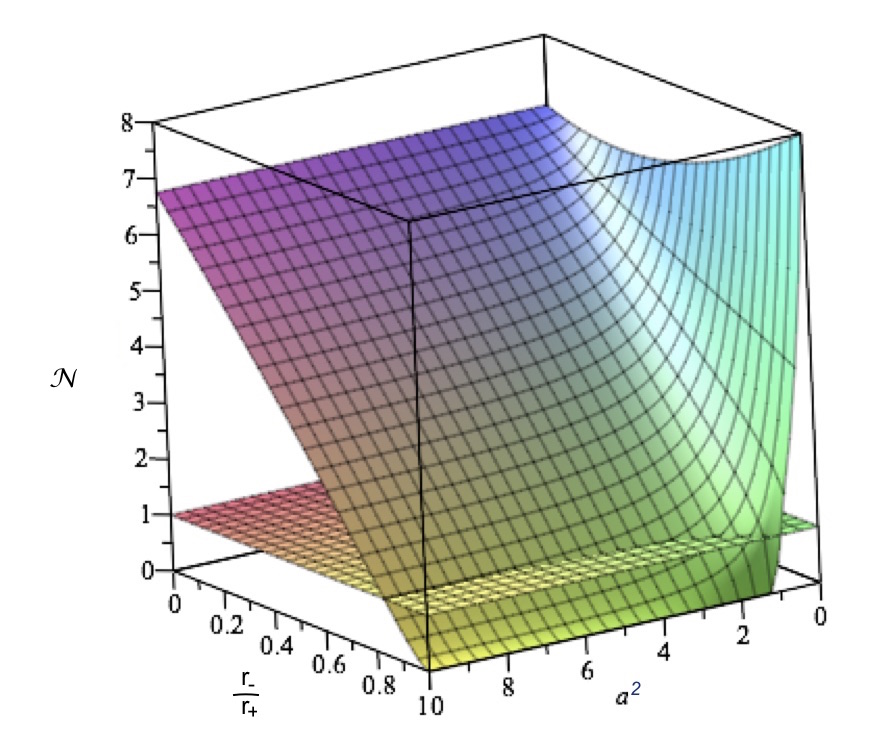}
\caption{
The ratio  ${\cal N}=\frac{R_\opt(\bar r)^2}{4M_{ADM}^2}$  as a function of 
$ \frac{r_-}{r_+}$ and  $a^2$ .\label{Nr}}
\end{center}
\end{figure}

For the dilatonic black holes with $\lambda =0$, it  is straightforward to show that this bound is  satisfied when 
$a^2\le 1$ for  any value of the ratio  $ \frac{r_- }{r_+}\le 1$.  At a critical value $a^2=1$, we have ${\cal N}= 1$ for  $\frac{r_-}{r_+}= 1$.  For $a^2>1$   the bound 
is violated, i.e., ${\cal N} <1$  for 
sufficiently large  values of the ratio  $\frac{r_- }{r_+}$. In the 
limiting case of  large $a^2$,  the bound is violated for  
$0.85 \lesssim \frac{r_- }{r_+}$.
These features are quantitatively displayed in
Figure 1, which  depicts  the value of 
${\cal N}$   as a function of $\frac{r_-}{r_+}$ and
$a^2$.
The figure further confirms that  ${\cal N}$ is  bounded from 
above by $8$, and that it saturates this bound for the extremal 
Reissner-Nordstr\"om black hole:
\be
R_\opt(\bar r)  \le 4\sqrt{2} \, M_{ADM}\,.
\ee
 This bound is saturated for the 
extremal Reissner-Nordstr\"om black hole.

Hod's theorem (\ref{hodtheorem}) states that
\ben
R(\bar r)=R_\opt(\bar r)\, (-g_{tt}(\bar r))^{\frac{1}{2}}\le 3M_{ADM}\, .
\een
This is  clearly satisfied, since both $R_\opt(\bar r)$ and 
$|g_{tt}(\bar r)|$ are bounded from below.
The bound is saturated when the ratio $\frac{r_-}{r_+}$  goes to zero.
We illustrate these results in Figure 2. 

\begin{figure}[ht]
\begin{center}
\includegraphics[height=8cm]{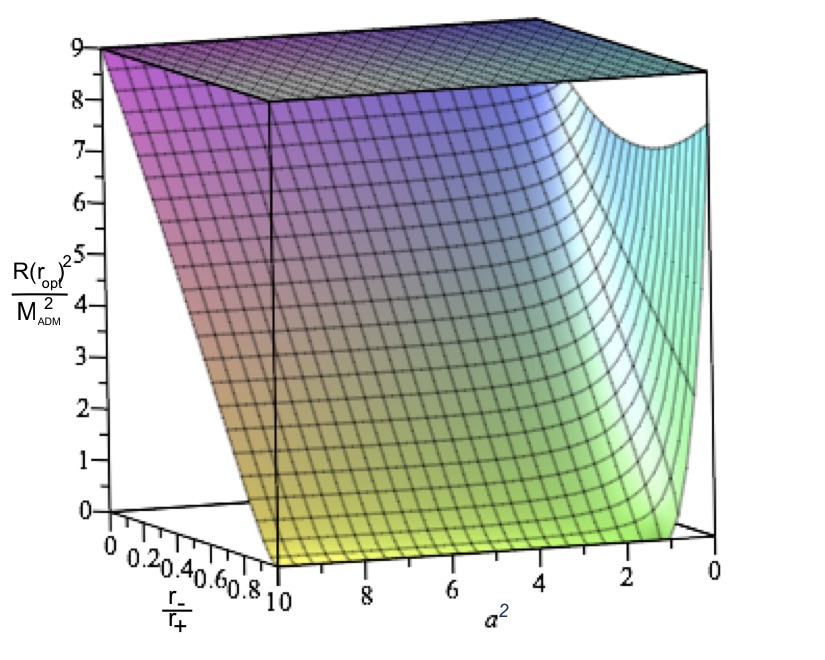}
\caption{ The ratio of $\frac{R(\bar r)^2}{M_{ADM}^2}$   
is plotted as a function of $\frac{r_-}{r_+}$  and $a^2$.  Note the 
ratio is always smaller than 9, thus confirming the bound. }
\end{center}
\end{figure}

\subsubsection{Photon spheres for  ultra-extremal dilatonic solutions}

We now turn to the analysis of photon spheres in the case when the solutions 
have a mass below the BPS bound, i.e.\ ultra-extremal black holes.  
It is convenient parameterise $r_{\pm}$ in terms of the charge and the ADM mass of the black holes:
\bea
r_+&=&M_{ADM}+\sqrt{M_{ADM}^2-(1-a^2)Q^2}\, , \nn\\
r_-&=& \frac{1-a^2}{1+a^2}\left(M_{ADM}-\sqrt{M_{ADM}^2-(1-a^2)Q^2}\right)\, .
\eea
 
 \begin{figure}[ht]
\begin{center}
\includegraphics[height=8cm]{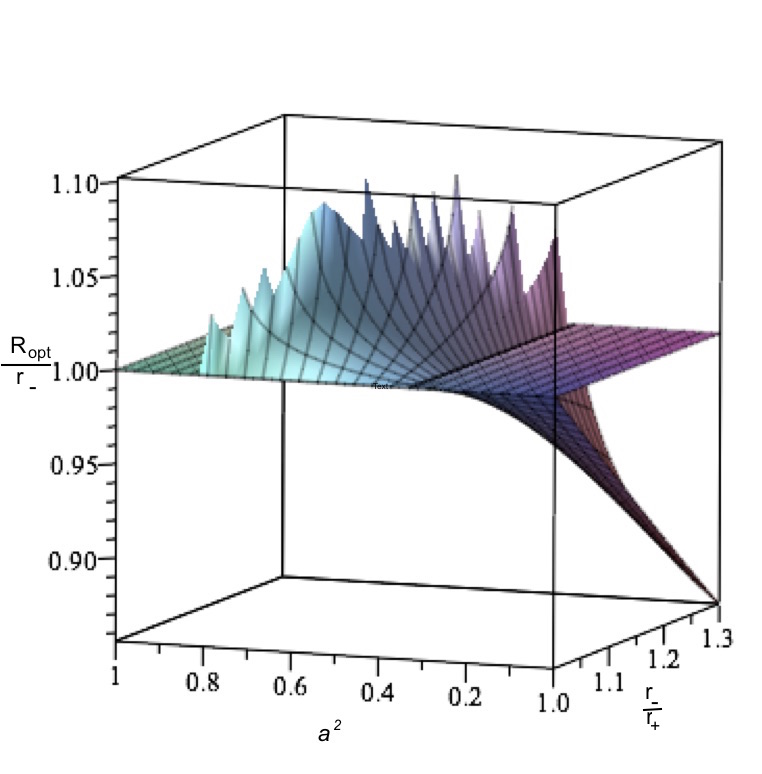}
\caption{ The range of the second extremal photon radius plotted for  
$\frac{R_{opt}(\bar r)}{r_-}$  as a function of $\frac{r_-}{r_+}\ge 1$  
(ultra extremal solutions) and $ a^2$. Note that for $a^2\ge \frac{1}{3}$, 
there is always a range of $\frac{r_-}{r_+}>1$ for which the  second 
extremal photon radius is larger than $r_-$, and thus outside the 
naked singularity. }
\end{center}
\end{figure}
 The extremal black hole with the property  $r_{+}=r_-$ saturates the 
BPS bound:
\ben
 M_{ADM}^2= \frac{Q^2}{1+a^2}\, ,
\een
Note that for $a^2\le 1$, there is a range  of ultra-extremal black
holes with
 \ben 
\frac{Q^2}{1+a^2}\ge M_{ADM}^2\ge (1-a^2) Q^2\,.
\een
In this regime, $\frac{r_-}{r_+}\ge 1$, namely, the outer horizon is 
at $r_-$ and the inner one at $r_+$.  From the analysis of the extremal 
equation of the photon sphere it is now possible to show that for 
$ \frac{1}{3} \le a^2\le 1$, {\it both}   extrema of the photon 
sphere (\ref{extremad})  lie  {\it outside} the larger horizon $r_-$, 
as long as
\ben
1\le \,\frac{r_-}{r_+} \,
\le \frac{9(a^2+1)}{3a^2+7+4\sqrt{2(3a^2-1)}}\,,\label{bounde}
\een
For $a^2$   in the range  $\{\frac{1}{3},1\}$,  the upper bound in 
(\ref{bounde})  has the range $\{\frac{3}{2},1\}$.  We illustrate these 
results in Figure 3.
  In this range of parameters the outer photon radius corresponds to 
a minimum, which is
stable (an anti-photon sphere), and the  
inner solution to a maximum, which is therefore unstable (a photon sphere).

\subsection{Projective symmetry for the dilatonic black holes}

Here we demonstrate that the static dilatonic black holes also exhibit 
the projective symmetry, just as we  demonstrated for the static STU black 
holes in Subsection \ref{projSTU}. 

The radial transformation that casts the metric in the form (\ref{one})
that makes the projective symmetry manifest 
can be integrated to give:
\be
u=\frac{1}{r_-}\frac{1+a^2}{1-a^2}\left(1-F_-^\frac{1-a^2}{1+a^2}\right)\, , 
\ee
with $F_{\pm}=1-\frac{r_{\pm}}{r}$. This equation can then be inverted, 
to give $r$ in terms of $u$.
We have already shown that 
\be
k(u)=\frac{1}{R_\opt^2}=\frac{1}{r^2}F_+F_-^\frac{1-3a^2}{1+a^2}-\lambda
\ee
has a  cosmological constant contribution that is independent of the 
radial coordinate.
The $a=0$ case is special, with
\be 
u=-\frac{1}{r_-}\log(1-\frac{r_-}{r})\, .
\ee

\section{Black Holes in Horndeski Gravity}\label{Horn}

   In this section we examine the static black hole solutions in a
simple example of a Horndeski theory of gravity coupled to a scalar 
field, and we show that in certain cases there can be two photon spheres 
outside the black hole horizon. Specifically, we consider the
theory described by the Lagrangian
\be
{\cal L}= \sqrt{-g}\,\Big[ R -2\Lambda - \ft12(\alpha\, g^{\mu\nu} -
  \gamma\, G^{\mu\nu})\, \del_\mu\chi\, \del_\nu\chi\Big]\,,
\ee
where $G_{\mu\nu}=R_{\mu\nu}-\ft12 R g_{\mu\nu}$ is the Einstein tensor.
In four dimensions, the black hole is given by \cite{anciol,felilupo}}
\bea
ds^2&=& -h dt^2 + \fft{dr^2}{f} + r^2\, d\Omega_2^2\,,\qquad
{\chi'}^2 = \fft{3\beta\, g^2\, r^2}{(1+3 g^2 \, r^2)\, f}\,,\nn\\
h&=& C-\fft{\mu}{r} + g^2\, r^2 + 
 \fft{D\, \arctan(\sqrt3 g r)}{\sqrt3 g r}\,,\quad
f=\fft{(4+\beta\gamma)^2 (1+3g^2\, r^2)^2}{[4+ 3(4+\beta\gamma) g^2\, r^2]^2}
\, h\,,
\eea
where
\be
C= 
 \fft{4-\beta\gamma}{4+\beta\gamma}\,,\qquad 
 D=\fft{\beta^2\gamma^2}{(4+\beta\gamma)^2}\,,
\ee
and the constants $g$ and $\beta$ are related to $\alpha$, $\gamma$ and 
$\Lambda$ by
\be
\alpha= 3 g^2\, \gamma\,,\qquad \Lambda=-3 g^2\, (1+\ft12 \beta\gamma)\,.
\ee

  Defining 
\be
G(x)\equiv \fft{\arctan x}{x}\,,
\ee
and letting $x=\sqrt3\, g r$, the horizon is located at $r=r_0$
(and hence $x=x_0$) where
\be
0= -\fft{\mu}{r_0} + C + (g^2 r_0^2 + D G(x_0))\,.
\ee
Now $3x^2 +G(x)-1\ge0$, and $D\le 1$, and so it follows that
\be
g^2 \, r^2 + DG(\sqrt3 g r) \ge (g^2\, r^2 + D G(\sqrt3 g r))\Big|_{g=0}\,,
\ee
and so the radius $r_0$ of the horizon for general $g$ is smaller than
the radius when $g=0$, implying
\be 
r_0\le \fft{\mu}{C+D} =\fft{16\mu}{(4+\beta\gamma)^2}\,.\label{horineq}
\ee

  The photon sphere is determined by finding the root or roots of
$(R^{-2})'=0$ that lie outside the horizon, where $R^2=r^2/h$ is the
radius-squared in the optical metric.  Note that unlike all the
previous black hole examples, here $(R^{-2})'$ is dependent on the 
``gauge coupling'' $g$ that determines the effective AdS cosmological
constant, since it enters in the function $G(\sqrt3 g r)$.  Setting
$(R^{-2})'=0$ we obtain an expression that can be written as
\be
1-\fft{3\mu}{2(C+D)\, r}= \fft{D}{2(C+D)}\, \Big[\fft{3+2x^2}{1+x^2} 
  -\fft{3 \arctan x}{x}\Big]\,.\label{photoncon}
\ee
The function in square brackets on the right-hand side can be shown to be
non-negative, and hence we have the result that the radius $r_s$ of the
photon sphere obeys the inequality
\be
r_s \ge \fft{3\mu}{2(C+D)}= \fft{24\mu}{(4+\beta\gamma)^2}\,.
\ee

In view of (\ref{horineq}), we see that the photon sphere must lie
outside the horizon, with
\be 
r_s \ge \ft32 r_0\,.
\ee

  We can write (\ref{photoncon}) as
\be
\fft{32}{\beta^2\gamma^2} - 
 \fft{3\sqrt3\, g\, \mu\, (4+\beta\gamma)^2}{\beta^2\gamma^2\, x} =
  \fft{3+2x^2}{1+x^2}
  -\fft{3 \arctan x}{x}\,,
\ee
and since the right-hand side ranges monotonically 
from 0 to 2 as $x$ ranges from 
0 to infinity, it follows that there 
will generically be two solutions or none if $32/(\beta^2\gamma^2)<2$ 
(depending on the value of $\mu$), and one solution or none if
$32/(\beta^2\gamma^2)>2$ (again, depending on the value of $\mu$). 

\section{Quintessence Black Holes}

According to \cite{Kiselev:2002dx}, quintessence should satisfy
\ben
 T_{\hat \phi \hat \phi} = T_ {\hat \theta \hat \theta} = -\half(3w+1)  
  T_{\hat r \hat r}= \half(3w+1) 
 T_{\hat t \hat t}   \,,
\een 
where $w$ is taken to be a constant.
The dominant energy condition \cite{Hawking:1970eq} 
requires  $T_{\hat t \hat t} \ge 0$ and 
\ben
|3 w +1| \le 2 \,.
\een
It follows from (\ref{dgamdr}) that $\gamma$ in the metric
(\ref{MSmetric}) is constant, and hence, by rescaling $t$ appropriately, 
\ben
-g_{tt} = \frac{1}{g_{RR}} =  \frac{1} {1-\frac{2M(R)} {R}  } \,,
\label{Kiselev} 
\een
where $R$ is the area distance. $M(R)$ is called  the Misner-Sharp  mass. 
For  further discussion of (\ref{Kiselev})  see \cite{Jacobson:2007tj}.
On then has 
\ben
\frac{2M(R)}{R} = \frac{2M_0}{R} +  \epsilon\bigl(\frac{L_w}{R} \bigr)^ {3w+1} \,.  
\een 
The values $(w,\epsilon) =(\frac{1}{3} , -1) $ 
corresponds to the Reissner-Nordstr\"om metric.   If $(w,\epsilon) =(-1, \pm 1)$, 
one has  a  cosmological constant.  Kiselev \cite{Kiselev:2002dx} 
favours, on symmetry grounds,  $(w, \epsilon)  =(-\frac{2}{3},1)$
for  quintessence which, as a consequence,
satisfies the dominant energy condition. Under this assumption, the metric
is given by 
 \ben 
ds^2 =- \Big( 1- \frac{2M}{R} - \frac{R}{L} \Big) dt ^2  + 
\frac{dr^2}{1- \frac{2M}{R} - \frac{R}{L} } + 
R^2 ( d \theta ^2 + \sin \theta ^2 
d \phi ^2 ) \,. \label{quint} 
\een
If $M=0$ we obtain  a metric reminiscent of de Sitter space, 
with a cosmological event horizon at $R=L$
and a naked singularity at $R=0$. The optical radius $R_\opt$  is given by
\be
\frac{1}{R_\opt^2} = \frac{1}{R^2} - \frac{1}{LR}\,, 
\een 
and so 
\ben\frac{d}{dR} \bigl(\frac{1}{R_\opt^2}  \bigr ) =  
-\frac{1}{R^3}  \bigl( 2 -  \frac{R}{L} \bigr ) \,,
\een
which is  negative throughout the static region. 

One may take $ L$ negative;  $L=-a$ say. 
This corresponds to quintessence with a negative energy density.  
The metric no longer has a cosmological horizon, but it does  
not have AdS asymptotics, but, rather, something softer.   
Defining 
\ben
\rho +a = a\sqrt{1+ \frac{R}{a}} \,,
\een
so that if
\ben 
R=r + \frac{r^2}{4 a}  \,,
\een 
the metric becomes
\ben
ds^2 = - dt ^2 + dr^2 + r^2 (1 + \frac{r}{4 a} )^2  ( d \theta ^2 + 
\sin \theta ^2 d \phi ^2 ) \,.
\een
In the positive $r$ direction, the  area of a sphere of constant radius 
increases faster than it would in flat space, but more slowly than in 
AdS$_4$. In the negative $r$ direction we get the solution
for ordinary quintessence with a cosmological horizon.
The solution has a singularity at $r=0$. This is clear,  since 
$R^2$ as a function of $r$ has odd powers  of $r$, 
starting with an  $r^3$ term.
   
We  turn now to the  quintessence black hole (\ref{quint}) with $M>0$.   
If $M <L/8 $  then there are two Killing horizons, at 
\ben
R= R_{H_\mp} = \half L \bigl ( 1 \mp \sqrt{1- \frac{8M}{L}}  \bigr )  
= \half L \bigl ( 1 \mp \sqrt{1- 8 x} \bigr )\,,
\een
where $x=M/L$.  These horizons coalesce at $R=\frac{L}{2}$ 
when $M =L/8$,  or $x=1/8$ . 

   Provided $M < L/6$, i.e. $x<1/6$, which of course is always greater  
than the critical value $x=1/8$,  the derivative  
\ben
\frac{d}{dR} \bigl(\frac{1}{R_\opt^2}  \bigr ) =  
\frac{1}{LR^4} \bigl (R^2 -2RL  + 6ML   \bigr)
\een
vanishes at 
\ben
R= \bar R_\mp  = L \bigl(1 \mp   \sqrt{1- \frac{6M}{L}}           
          \bigr ) =  L \bigl(1 \mp   \sqrt{1-6x}) \,.  
\een
Now  $-g_{tt}$ vanishes at the horizons $R=R_{H_\mp}$. Thus we expect 
an odd number of critical points
in the static interval $R_{H_-} < R< R_{H_+} $.  Since we have two solutions, 
we therefore expect that one will   
lie  inside the static region and one outside. In order to see which 
we calculate  
\bea
\bar R_- -  R _{H_-} &=& \frac{L}{2}  (  1 -f(x) )\\ 
\bar R_+  - R_{H_+} &=& \frac{L}{2}  ( 1 +  f(x)  )\,,
\eea
where  the function  
\ben
f(x):= 2 \sqrt{ 1-6x } -\sqrt{1-8x}
\een
is defined on the  interval   $0\le x \le \frac{1}{8}$.  Clearly
\ben
f(0) = f(\frac{1}{8}) = 1\,,\qquad  
f^\prime(x) = -\frac{6}{\sqrt{1-6x}} +  \frac{4}{\sqrt{1-8x}} \,. 
\een
Any critical point of $f(x)$ must  satisfy 
\ben
9 (1-8x) = 4 (1-6x) 
\een 
There is a unique such $x$, namely
\ben
x= \frac{5}{48}\,,\qquad f(\frac{5}{48})  = \sqrt{\frac{2}{3}}  \,,
\een
and hence 
\ben
\sqrt{\frac{2}{3}} \le  f(x) \le 1\,,  
\een
and so
\ben
1 \pm f(x) \ge 0\,.
\een
Thus 
\ben
R_{H_-} \le \bar R_-  \le R_{H_+} \le \bar R_+ \,. 
\een
Hence we obtain a single photon sphere, with the larger critical point 
lying beyond the cosmological horizon.  There is no anti-photon sphere.

\section{Higher Dimensions}

\subsection{Five dimensions} 

 The metric of the static three-charge black hole solution of the 
maximally supersymmetric 
gauged supergravity \cite{Behrndt,CG} takes the form 
\ben
ds^2 = -(H_1H_2H_3)^{-2/3} f dt^2  + (H_1H_2H_3)^{1/3} 
\Bigl ( f^{-1} dr^2 + r^2 d \Omega _3^2 \Bigr )  \,, \label{five}
\een
where
\ben
f = 1 - \frac{2m}{r^2}  + g^2r^2 H_1H_2H_3 \,\qquad  H_i = 1 +
\frac{q_i}{r^2} \, ,\quad i = 1, 2, 3\, .  
\een
The mass and  three U(1) charges are given by:
\ben
M_{ADM}=m+\frac{1}{3}\sum_{i=1}^3 q_i\, , \quad \quad
Q_i^2=q_i(q_i+2m)\, , \quad \quad i=1,2,3\, .
\een
Using (\ref{five}, 
  we see that the three-charge black hole in AdS$_5$  has an 
optical radius $R_\opt (r)$ given by
\ben
\fft{1}{R_\opt ^2} = {1 \over r^2 H_1H_2H_3  }
\Big(1-{2m  \over r^2}  +g^2 r^2 H_1H_2H_3 \Big) =  
{1 \over r^2 H_1H_2H_3  }
\Big(1-{2m  \over r^2} \Big)  +g^2  \,.
\een  

The situation is very similar to that in four spacetime dimensions.
The extremum is determined by the equation
\ben
\frac{r^2-4m}{r^2-2m}=\sum_{i=1}^3\frac{q_i}{r^2+q_i}\,,\label{ext5}
\een
which has a unique positive solution with  $r^2=\bar r^2 >4m$.

 A generalization Hod's theorem (\ref{hodtheorem}) to higher dimensions, given  in \cite{gallo}, can be shown to be satisfied for these solutions.
Namely, one can write
\ben
\bar R^2 =\prod_{i=1}^3 ({\bar r}^2+q_i)^{\frac{1}{3}}\le 
\frac{1}{3}\sum_{i=1}^3({\bar r}^2 +q_i)\le  \frac{1}{3}\sum_{i=1}^3(4m+q_i)
= 4M_{ADM}+{\bar r}^2-4m-\sum_{i=1}^3 q_i \le 4M_{ADM}\, .
\een
The first inequality above is due to the inequality of geometric and 
arithmetic means, and the second inequality follows from:
\ben
{\bar r}^2-4m-\sum_{i=1}^3 q_i=-\sum_{i=1}^3 \frac{q_i(q_i+2m)}{{\bar r}^2 +q_i}\le 0\, ,
\een
where the first equality above  is due to  (\ref{ext5}).
 
\subsection{Seven dimensions}

The static two-charged black hole in an AdS$_7$ background given in \cite{CG}
has the metric
\ben
 -(H_1H_2)^{-{4 \over 5}}  f dt ^2 + (H_1H_2) ^{1 \over 5}
\Bigl ( { dr ^2 \over f} + r^2 d \Omega ^2 _ 5 \Bigr ) \,,
\een
with 
\ben
f= 1 -{2m \over r^4} + g^2 r^2 H_1H_2 \,,\qquad H_i= 1+ {q_i  \over r^4 } \, , \quad i=1,2\, .
\een
The mass and  two U(1) charges are given by:
\ben
M_{ADM}=m+\frac{2}{5}\sum_{i=1}^2q_i\, , \quad \quad
Q_i^2=q_i(q_i+2m)\, , \quad  i=1,2\, .
\een
The optical radius $R_\opt(r)$ is given by
\ben
{1 \over R_\opt^2 }
 = {1 \over r^2 H_1 H_2} \Big(1 -{2m \over r^4} + g^2 r^2 H_1H_2\Big)  
=  {1  \over r^2 H_1 H_2} \Big(  1 -{2m \over r^4} \Big)  + g^2\,.
\een
and the  argument goes through as in the previous example. 
The extremum is determined by the equation
\ben
\frac{r^4-6m}{r^4-2m}=\sum_{i=1}^2\frac{2q_i}{r^4+q_i}\,,\label{ext7}
\een
which has a unique positive solution with $r^4=\bar r^4 >6m$.
 
 It can be shown that these solutions satify an analog of  Hod's theorem (\ref{hodtheorem}), generalised to seven dimensions \cite{gallo}.
Namely, we write
\ben
\bar R^4 =[{\bar r}^4\prod_{i=1}^2 ({\bar r}^4+q_i)^2]^{\frac{1}{5}}\le 
\frac{1}{5}[{\bar r}^4 +2({\bar r}^4+q_1)+2({\bar r}^4+q_2)]\le 
6M_{ADM}+ {\bar r}^4-6m-2\sum_{i=1}^2q_i \le 6M_{ADM}\, .
\een
The first inequality above is due to the inequality of geometric and 
arithmetic means. The second inequality is due to:
\ben
 {\bar r}^4-6m-2\sum_{i=1}^2q_i =-2\sum_{i=1}^4\frac{q_i(q_i+2m)}{{\bar r}^4+q_i}\le 0
\, ,
\een
where the first equality above is due to (\ref{ext7}).
 
\section{Conclusions}

In this paper we have examined  the optical metrics
of  static spherically  symmetric solutions
of various theories of current interest. In particular we have been
interested  in whether the they admit photon spheres and if so how many.
In the case of all the solutions we have looked at whose  energy momentum tensor
satisfies  the dominant and strong energy conditions and which 
are non-singular outside a  regular event horizon we have fond a unique 
photon sphere and as a consequence no anti-photon spheres.
For some ultra-extremal solutions we have found, consistent with other
authors one may have both a photon sphere and an anti-photon sphere.
We have also found in the case of a particular theory of Horndeski
type that one may have both a  photon sphere and an anti-photon sphere
outside a regular Killing horizon of the spacetime metric.
We are thus lead to the conjecture that a violation of
the  either the dominant or the strong energy condition
is a necessary condition for the existence  of an anti-photon sphere
outside a regular black hole horizon.
       
We have investigated a conjecture of Hod  \cite{Hod2}, concerning a
lower bound on the
optical radius of the photon sphere (see eqn (\ref{Hod})), 
and found counterexamples in the case of static black holes in STU supergravity
where fewer than three electric charges are turned on.   
 
We have also  found that that the rather mysterious
projective symmetry of the optical metric first
observed in the case of the Schwarzschild de Sitter metric
continues to hold for the static spherically symmetric solutions
of the STU supergravity theories. At present we have no conceptual
understanding  of why this symmetry is present, nor why it seems 
related to the fact that the null geodesics in this case 
may be described by Weierstrass elliptic functions.

\section{Acknowledgements}

We are grateful to Emanuel Gallo, 
Sharhar Hod and Claude Warnick for some helpful remarks.
The work of M.C. is supported in part by the DOE (HEP) Award DE-SC0013528, 
the Fay R. and Eugene L. Langberg Endowed Chair (M.C.) and the Slovenian 
Research Agency (ARRS).  The work of C.N.P. is supported in part by DOE 
grant DE-FG02-13ER42020. 
G.W.G. is grateful for the  award of a LE STUDIUM Chair held at the  LMPT
of the University of Tours
under the auspices of which some of the work described in this paper
was carried out. M.C. and C.N.P. are grateful the University of Tours
and Beijing Normal University for
hospitality during the  course of the work.

\appendix

\section{ k-Essence and  Irrotational Relativistic Fluids }

The equation of motion for the theory with Lagrangian $L=L(X)$, where 
$X=-g^{\mu \nu}\p_\mu  \psi  \p_\nu \psi$, is given by
\ben
\nabla _\mu  \Bigl(\frac{\p L}{\p X} \nabla ^\mu \psi\Bigr ) =0\,.
\een
We may define a   current 
\ben
J^\mu =  \frac{\p L}{\p X} \nabla ^\mu \psi\,,
\een
which is conserved by virtue of the shift symmetry
$\psi \rightarrow \psi +{\rm constant}$.
If  $L_X = \frac{\p L}{\p X}$, then the energy-momentum tensor
is 
\ben
T_{\mu \nu} =2 L_X \p_\mu \psi \p_\nu \psi  + g_{\mu \nu} L 
\een
If $X>0$ we may define a unit timelike vector by
\ben
u_\mu = \frac{\p_\mu \psi}{\sqrt{X}}\,,
\een
and find that the energy-momentum tensor takes the form
of an irrotational perfect fluid  with Eulerian 4-velocity
$u_\mu$ :
\ben
T_{\mu \nu} =  \rho u_\mu u_\nu + P ( g_{\mu \nu} +u_\mu u_\nu )\,,
\een
where 
\ben
\rho + P = 2XL_X \,,\qquad P = L \,,\qquad \rho=2XL_X-L  \,.
\een
Here $g_{\mu \nu} +u_\mu u_\nu = h_{\mu \nu }$ is a a projection tensor
which projects an arbitrary vector to one  orthogonal to the world lines
of the fluid. A simple calculation yields 
\ben
\frac{\p \rho}{\p P}= \frac{L_X -2X L_{XX} }{L_X}\,,  \label{identify}
\een
whence, as will be verified  later,  the sound speed $v_s$ is given by
\ben
v_s^2 = \frac{L_X}{L_X -2 X L_{XX} }\,. \label{sound}
\een 
Examples of k-essence include
\begin{itemize} 
\item 
Polytropic fluid with $P=w\rho$ 
\ben
L= X^{\frac{1+w}{2w}} = X^p \,, 
\qquad w={\rm constant}=\frac{1}{2p-1}\,,
\een
where $p$ may be fractional. The left-hand side  of the equation of motion
\ben
\nabla ^\mu \bigl(X \nabla _\mu \psi  \bigr ) =0 
\een
is what  one might call  $p$-D'Alembertian, the analogue
in Lorentzian  geometry of the
$p$-Laplacian of Riemannian geometry.
The case $p=2$ in $d=4$ is conformally  invariant. 
\item 
Born Infeld: 
\ben
L= -\sqrt{1-X} +1 \,,\qquad P=\frac{\rho}{\rho+1}  \,.
\een
\item
The Chaplygin gas: 
\ben 
L= -\sqrt{1-X} \,,\qquad P=-\frac{1}{\rho} \,.
\een
\end{itemize}

Of course the fluid description only works if $X>0$ and so, in particular,
it cannot be applied to  static solutions, which have $X<0$.  

\subsection{Thermodynamics}

Since $u^\mu _{;\mu} = \dot V/V$, where $V$ is the infinitesimal
volume of an  element of the fluid dragged along the flow lines,
the first law of thermodynamics reads 
\ben
(\rho +P  )  dV +  V d\rho   =0\,.  \label{thermo2}
\een
Now in general, if a   fluid is locally homogeneous  and  passes through 
thermodynamic equilibria, we have 
\ben
Ts=\rho+P \,, \qquad  Tds = d \rho \,, 
\qquad \frac{d \rho}{\rho+P} = \frac{ds}{s} \,.    
\label{duhem}
\een
Therefore,  by (\ref{thermo2}), we have
\ben
s V = {\rm constant}
\een
and the flow is isentropic.
From (\ref{duhem}) the dependence of 
all $(\rho,P,s,T)$ on any one of them is determined once an equation 
of state
is specified, and hence by (\ref{thermo2}) on the volume expansion.
Thus for  a polytrope, 
\ben
\rho= A \Big(\frac{T}{1+w}  \Big)^{\ft{1+w}{w}} \,,\qquad s=
(1+w )A  \Big(\frac{T}{1+w}  \Big)^{\ft{1}{w}} \,,
\een
where $A$ is a constant with dimensions $L^{-3}\, M^{-\ft{1}{w}}$. 
If $w=\frac{1}{3}$,  $A$ has dimensions $L^{-3}\, M^{-3} =\hbar^{-3}$.
If $w \ne\frac{1}{3}$ one needs a further dimensionful constant 
to relate the energy density to the entropy density  or to the temperature. 

\subsection{Entropy current as Noether current}

The conserved  current arising from the shift symmetry
$\psi \rightarrow \psi +{\rm constant}$ 
gives rise to a conserved current,  
\ben
J^\mu = {\p L \over \p ( \p_\mu \psi ) } = - 
2 X^\half L_X  \,   u^\mu \,.
\een 
From (\ref{identify}) 
\ben
- 2X^\half L_X  = -  X ^{-\half} 2 X L_X =- (\rho +P) X^{-\half} \,,     
\een
and from (\ref{duhem}) we have 
\bea
\frac{ds}{s}  &=&   \bigl ( \frac{ d \rho + dP}{\rho+P} - \frac{dP}{\rho + P}  
 \bigr ) 
\\
& =&  d \ln ( \rho + P) - \frac{dL} {2X\, L_X}\\    
& =&   d  \ln ( \rho + P)  -  \frac{dX}{2 X}\,,   
\eea
whence 
\ben
s = {\rm constant}\times  (\rho+P ) X^{-\half} \,.
\een
Thus
\ben
J^\mu = {\rm constant} \times  s u^\mu \,.
\een
For example, for radiation we have $ w= \frac{1}{3}$, and hence  
\ben
L=X^2= ( g^{\mu \nu} \del_\mu\psi \del_\nu \psi)^2\,. 
\een
The equation of motion is
\ben
\nabla _\mu \bigl 
( ( \nabla \psi)^2 \nabla^\mu \psi \bigr) =0\,, \label{radeom2}  
\een
or, as long as $\nabla ^\mu  \psi$ is timelike,
\ben
\bigl ( g^{\mu \nu} - 2 u^\mu u^\nu \bigr) \nabla _\mu \nabla _\nu \psi =0\,. 
\een
One recognizes
\ben
( a^{-1})^{\mu \nu} =  g^{\mu \nu}-2 u^\mu u^\nu 
\een  
as the acoustic  co-metric, i.e.\ the inverse 
of the acoustic metric
\ben
a_{\mu \nu}= g_{\mu \nu} +{2 \over 3} u_\mu u_\nu 
\een
for a fluid with $P={1 \over 3} \rho $.

If one repeats  the  calculation above for 
$L=X^p$, one finds 
\bea
( a^{-1})^{\mu \nu} &=&  g^{\mu \nu}-(2p-1)  u^\mu u^\nu\\ 
a_{\mu \nu}&=& g_{\mu \nu} +{1-w} u_\mu u_\nu \,,
\eea
which corresponds to a fluid with sound speed
$v_s=\sqrt{\frac{\p P }{\p \rho}} = \sqrt{w}$. 
For both the Born-Infeld and the Chaplygin gases,
 one finds the sound speed $v_s$  to be given by
\ben
v_s = \sqrt{1-X} 
\een
 and 
\bea
( a^{-1})^{\mu \nu} &=&  g^{\mu \nu}- \frac{X}{1-X}   u^\mu u^\nu\\ 
a_{\mu \nu}&=& g_{\mu \nu} +X  u_\mu u_\nu \,.
\eea

    In general one finds that the equation of motion for $\psi$ takes the form
\ben
({a^{-1}})^{\mu \nu} \nabla _\mu \nabla _\nu \psi =0\,,
\een
where the acoustic co-metric  ${a^{-1}} ^{\mu \nu}$ is given by
\ben
({a^{-1}})^{\mu \nu} =g^{\mu \nu} -2\frac{L_{XX}}{L_X} u^\mu u^\nu \,.
\label{acoustic}
\een
Equation (\ref{acoustic}) is consistent with (\ref{sound}):
\ben
\nabla _\mu \bigl 
( ( \nabla \psi) ^2    \nabla ^\mu \psi \bigr ) =0\,, \label{radeom}  
\een
or, as long as $\nabla ^\mu  \psi$ is timelike, 
\ben
\bigl ( g^{\mu \nu} - 2 u^\mu u^\nu \bigr) \nabla _\mu \nabla _\nu \psi =0\,. 
\een

\subsection{Black hole accretion and emission}

In order to describe a steady (i.e. time independent)  
spherically symmetric
flow in a background whose metric is 
\bea
ds ^2 &=& -\Delta (R)  dt ^2 + \frac{dR ^2 }{F(r)}
+ R^2  \bigl( d \theta ^2 + \sin ^2 \theta d \phi ^2 \bigr ) 
\\ &=& \Delta \Bigl \{-dt ^2 + {dr_\star } ^2 + {r^2 \over \Delta}   
 \bigl( d \theta ^2 + \sin ^2 \theta d \phi ^2 \bigr ) \Bigr \}
\eea
where the metric in the braces is the optical metric
and $r_\star$ is the radial optical distance, often called the
the Regge-Wheeler tortoise coordinate:
\ben
d r_\star = \frac{ d R}{ \sqrt{F  \Delta}} \,.
\een
We make the ansatz 
\ben
\psi = t -\chi(R)\,, \label{ansatz}
\een
and find that  the fluid 3-velocity $v$  with respect
to a local orthonormal frame at rest with respect to the hole
is given by
\ben
v={d \chi \over d r_\star} \,. 
\een 
If $v>0$, the flow is an outward-directed wind. If $v<0$, we have an  
inward-directed accretion  flow. Moreover
\ben
X= \frac{1}{\Delta (1-v^2)} \,.
\een
For any  steady radial conserved current we have
\ben
R^2 \sqrt{\frac{\Delta}{F}} J_R = {\rm constant} \,.
\een 
In our case, if $d=4$,  that means 
\ben
v R^2  \,  L_X(X)  = {\rm constant} = v R^2\, L_X\,\Big( \frac{1-v^2}{\Delta}
\Big)   \,. 
\een
For a polytropic gas this gives 
\ben
v(1-v^2) ^{p-1} = a^2 \frac{\Delta ^{p-1}}{R^2}    
\label{eqn} \een
where $a$ is a constant. 
As $a$ varies, we obtain a family of curves in the $(v,r)$ plane,
labelled by the constant $a$. 
In the asymptotically-flat case, 
we are looking either for an  ingoing curve or an outgoing curve.   

It is  a simple matter to check that  (\ref{eqn})  with $p=2$ reproduces
equation (15) of \cite{Carter}. In the Schwarzschild  case 
\ben
\Delta = F=  1-{2 M \over R} \,,
\een
one finds that if  $R$ is plotted against
$v$ for different values of 
the constant $a$,  one obtains  Figure 1 of \cite{Carter}. 
The left-hand side of (\ref{eqn}) with $p=2$ achieves its 
greatest (least) value of $\pm {2\over \sqrt{27}}$ at
 $v={d \chi \over d r^\star}= \pm {1 \over \sqrt{3}}$.
In other words the fluid velocity coincides with the velocity
of sound.
The right-hand side of (\ref{eqn}) achieves its 
greatest (least) value when the optical radius 
\ben
R_\opt = {R \over \sqrt \Delta}
\een
is stationary: In other words, at radii for which
there are circular null geodesics.
In order that $v$ be a single-valued  
function of $r$  
on the interval $r \in (2M,\infty)$,
we must therefore choose
\bea
 {\rm constant }&=& \pm {2  \sqrt{27}}  M^2\,,\\
v(1-v^2 )&=&  \pm {2 \sqrt{27}}  M^2 {\Delta \over R^2 } \,.
\eea
The Bondi radius, at which the two flows, one inward (-)  and one 
outward (+),
make a transition from
subsonic to supersonic, occurs at the photon sphere  $R=3M$.

If the constant is positive we have a wind, whilst if the constant is
negative we have accretion. 
Asymptotically we have
\bea
{\rm wind}\,(+) : &R \rightarrow\infty& \qquad v = 1-{\sqrt{27}}
\bigl({M \over R}  \bigr ) ^2 + \dots \,, \\  
&r \rightarrow 2M&  \qquad v = { \sqrt{27} (R-2M)   \over 4  M} +\dots \,,\\   
{\rm accretion}\,(-)  :&R\rightarrow\infty& \qquad 
v =  - {2  \sqrt{27} } 
\bigl({M \over R}  \bigr ) ^2 +\dots \,, \\  
&r \rightarrow 2M& \qquad v = -1+ { \sqrt{27} (R-2M)  \over 8  M} +\dots  \,.  
\eea

Near the acoustic horizon we have
\bea
{\rm wind}\,(+) : &R \rightarrow 3M &v= {1 \over \sqrt{3}} + \sqrt{2 \over 27} \bigl(
{R-3M \over M}  \bigr) +\dots\,,\\
{\rm accretion}\,(-)  :& R \rightarrow 3M & v= - {1 \over \sqrt{3}} + \sqrt{2 \over 27} \bigl(
{R-3M \over M}  \bigr) +\dots\,.
\eea

The case for  general $p$ is similar.
The left-hand side of (\ref{eqn}) achieves its maximum  for $v^2 = w$.  
The right-hand side reaches its maximum for 
\ben
r_{\rm Bondi} = \half M (3 + \frac{1}{w} ) \,. 
\een

The analogy that is often
made is with a de Laval nozzle. The throat or waist of the hourglass-shaped 
nozzle is a sonic
horizon, at which the speed of sound and the speed of the fluid coincide.
In the present case, this throat is the waist at $R=3M$ of the optical
wormhole whose geometry interpolates between flat space
as $r_\star \rightarrow + \infty$ to the event horizon at $r_\star
\rightarrow - \infty$, where the geometry approaches that  near the
conformal infinity of  hyperbolic
three-space \cite{GW} and whose radius
 curvature is given by the surface gravity, or $2 \pi$ times
the Hawking temperature. As pointed out in \cite{GW}, this behaviour
is universal for all black holes, and now we see that equally universal
is the fact the the sonic horizon coincides (for a radiation gas)  with
the  photon sphere.

\end{document}